\documentclass[11pt,english]{article}
\usepackage{lmodern}
\usepackage[T1]{fontenc}
\usepackage[latin9]{inputenc}
\usepackage{xcolor}
\usepackage{amsmath}
\usepackage{amssymb}
\usepackage{cancel}
\usepackage{stackrel}
\usepackage{graphicx}
\usepackage{setspace}
\usepackage{wasysym}
\usepackage{colortbl}
\usepackage{multirow}
\makeatletter

\usepackage{jheppub}

\usepackage{longtable}
\usepackage[font=small,labelfont=sc]{caption}
\usepackage[]{subcaption}
\usepackage{float}
\usepackage[compat=1.1.0]{tikz-feynman}
\tikzfeynmanset{warn luatex=false}
\usepackage[normalem]{ulem}
\usepackage{geometry}
\usepackage{booktabs}
\hypersetup{
linkcolor=teal,
urlcolor=teal,
citecolor=teal,
linktocpage=true
}

\definecolor{darkred}{rgb}{0.6,0,0}
\definecolor{darkpurple}{rgb}{0.5,0,0.5}

\def\z2{$\mathbb{Z}_2$}
\def\z3{$\mathbb{Z}_3$}
\def\321{$SU(3)_c \times SU(2)_L \times U(1)_Y$}
\def\one{\ensuremath{\mathbf{1}}}
\def\two{\ensuremath{\mathbf{2}}}
\def\three{\ensuremath{\mathbf{3}}}
\def\threeS{\ensuremath{\mathbf{\bar 3}}}
\def\GGM{\ensuremath{SU(5)}}
\def\555{\ensuremath{SU(5)^3}}
\def\five{\ensuremath{\mathbf{5}}}
\def\fiveS{\ensuremath{\mathbf{\bar 5}}}
\def\eight{\ensuremath{\mathbf{8}}}
\def\ten{\ensuremath{\mathbf{10}}}
\def\tenS{\ensuremath{\mathbf{\overline{10}}}}
\def\fifteen{\ensuremath{\mathbf{15}}}
\def\fifteenS{\ensuremath{\mathbf{\overline{15}}}}
\def\24{\ensuremath{\mathbf{24}}}
\def\tfive{\ensuremath{\mathbf{35}}}
\def\tfiveS{\ensuremath{\mathbf{\overline{35}}}}
\def\forty{\ensuremath{\mathbf{40}}}
\def\fortyS{\ensuremath{\mathbf{\overline{40}}}}
\def\ffive{\ensuremath{\mathbf{45}}}
\def\ffiveS{\ensuremath{\mathbf{\overline{45}}}}
\def\SU15#1{\ensuremath{\mathbf{#1}}}

\definecolor{ForestGreen}{RGB}{34,139,34}

\providecommand{\tabularnewline}{\\}

\title{\boldmath Tri-unification: a separate $SU(5)$ for each fermion family}

\author[a,b]{Mario Fern\'andez Navarro,}
\author[a]{Stephen F. King}
\author[c,d]{and Avelino Vicente}
\affiliation[a]{School of Physics \& Astronomy, University of Southampton, Southampton SO17 1BJ, UK}
\affiliation[b]{School of Physics \& Astronomy, University of Glasgow, Glasgow G12 8QQ, UK}
\affiliation[c]{Instituto de F\'isica Corpuscular, CSIC-Universitat de Val\`encia, 46980 Paterna, Spain}
\affiliation[d]{Departament de F\'isica Te\'orica, Universitat de Val\`encia, 46100 Burjassot, Spain}

\emailAdd{Mario.FernandezNavarro@glasgow.ac.uk}
\emailAdd{S.F.King@soton.ac.uk}
\emailAdd{avelino.vicente@ific.uv.es}

\abstract{
In this paper we discuss 
$SU(5)^{3}$ with cyclic symmetry as a possible grand unified theory (GUT). 
The basic idea of such a tri-unification is that there is
a separate $SU(5)$ for each fermion family, with the light Higgs doublet(s) arising from the third family $SU(5)$, providing a basis for charged fermion mass hierarchies. $SU(5)^{3}$ tri-unification reconciles the idea of gauge non-universality with the idea of gauge coupling unification, opening the possibility to build consistent non-universal descriptions of Nature that are valid all the way up to the scale of grand unification.
As a concrete example, we propose a grand unified embedding of the tri-hypercharge model $U(1)_Y^3$ based on an $SU(5)^{3}$ framework
with cyclic symmetry. 
We discuss a minimal tri-hypercharge example which can account for all the quark and lepton (including neutrino) masses and mixing parameters. We show that it is possible to unify the many gauge couplings into a single gauge coupling associated with the cyclic $SU(5)^{3}$ gauge group, by assuming minimal multiplet splitting, together with a set of relatively light colour octet scalars. We also study proton decay in this example, and present the predictions for the proton lifetime in the dominant $e^+\pi^0$ channel.
}

\makeatother

\usepackage{babel}
\begin{document}
\makeatletter
\gdef\@fpheader{}
\makeatother

\maketitle \flushbottom

\clearpage
\pagenumbering{arabic}
\setcounter{page}{1}
\allowdisplaybreaks

\section{Introduction}

The flavour problem remains one of the most intriguing puzzles of the Standard Model (SM), being responsible for most of its parameters. The origin of three families, which are identical under the SM gauge group, but differ greatly in mass, with the quark mixing being small while the lepton mixing is large, is not addressed, while the origin of CP-violation only adds to the mystery. It is quite common to address these puzzles by assuming that the fermions are distinguished by a new spontaneously broken family symmetry, however this is not the only way forwards.

Recently two of us proposed an embedding of the SM based on the existence
of one local weak hypercharge associated to each fermion family \cite{FernandezNavarro:2023rhv},
\begin{equation}
SU(3)_{c}\times SU(2)_{L}\times U(1)_{Y_{1}}\times U(1)_{Y_{2}}\times U(1)_{Y_{3}}\,, \label{eq:tri-hypercharge}
\end{equation}
where each SM fermion family $i=1,2,3$ is charged only under their
corresponding $U(1)_{Y_{i}}$ factor. Such a framework avoids the
family replication of the SM and is naturally anomaly-free. If the
Higgs doublet(s) only carry third family hypercharge, then the third
family is naturally heavier and the light families are massless
in first approximation, providing a novel way of addressing the flavour problem.

In this paper we propose a grand unified embedding of the tri-hypercharge \cite{FernandezNavarro:2023rhv} model based on a non-supersymmetric $SU(5)^{3}$ framework
with cyclic symmetry. This is a generalisation of $SU(5)$ grand unification \cite{Georgi:1974sy} in which we assign a separate $SU(5)$ group to each fermion family, together with a cyclic symmetry to ensure gauge coupling unification\footnote{For related ideas see Refs.~\cite{Salam:1979p,Rajpoot:1980ib,Georgi:1981gj,Barbieri:1994cx,Chou:1998pra,Asaka:2004ry,Babu:2007mb}}.
We discuss a minimal example which can account for all the quark and lepton (including neutrino) masses and mixing parameters. We show that it is possible to unify the gauge couplings into a single gauge coupling associated with the cyclic $SU(5)^{3}$ gauge group, by assuming minimalistic multiplet splitting, together with a set of relatively light colour octet scalars. We also study proton decay in this example, and present the predictions of the proton lifetime in the dominant $e^+\pi^0$ channel.

More generally, the $SU(5)^{3}$
framework proposed here may embed a broader class of gauge
non-universal models, reconciling the ideas of gauge non-universality with gauge coupling unification at the GUT scale.
In particular, $SU(5)^{3}$ may embed theories based on the family decomposition of the SM group,
such as the $SU(2)_{L}^{3}$ model \cite{Li:1981nk,Ma:1987ds,Ma:1988dn,Li:1992fi,Muller:1996dj,Chiang:2009kb}, the $SU(3)_{c}^{3}$ model \cite{Carone:1995ge}  or the aforementioned tri-hypercharge model,
as an alternative to the existing ultraviolet (UV) completions which are all based
on (variations of) the Pati-Salam (PS) group \cite{Bordone:2017bld,Greljo:2018tuh,Allwicher:2020esa,Fuentes-Martin:2020pww,Fuentes-Martin:2022xnb,Davighi:2022bqf,Davighi:2022fer,Davighi:2023iks}.
We note however that while most of the previous papers explain the
origin of the flavour structure of the SM, none of them provides a gauge
unified framework. In general, gauge non-universal models can address
the flavour puzzle at the price of complicating the gauge sector,
which in such theories may contain up to nine
arbitrary gauge couplings in the UV.

The layout of the remainder of the paper is as follows.
In section~\ref{sec:General-framework} we discuss a general $SU(5)^{3}$ framework for model building.
In rather lengthy section~\ref{sec:The-Tri-Hypercharge-model} we analyse an example $SU(5)^{3}$ unification model breaking to tri-hypercharge, including the charged fermion mass hierarchies and quark mixing, neutrino masses and mixing, gauge coupling unification and proton decay.
Section \ref{sec:conclusions} concludes the paper.
In Appendix~\ref{sec:app} we detail the energy regimes, symmetries and particle content of the considered example.
In Appendix~\ref{sec:Hyperons}
we tabulate all possible hyperon embeddings in $SU(5)^{3}$ representations with dimension up to $\mathbf{45}$.

\section{General \texorpdfstring{$\boldsymbol{SU(5)^{3}}$}{tri-unification} framework for model building\label{sec:General-framework}}

The basic idea is to embed the SM gauge group into a semi-simple gauge group containing three $SU(5)$
factors,
\begin{table}[t]
\centering{}%
\begin{tabular}{cccc}
\toprule 
\textbf{Field} & $\boldsymbol{SU(5)_1}$ & $\boldsymbol{SU(5)_2}$ & $\boldsymbol{SU(5)_3}$\tabularnewline
\midrule 
$F_{1}$ & \fiveS & \one & \one\tabularnewline
$F_{2}$ & \one & \fiveS & \one\tabularnewline
$F_{3}$ & \one & \one & \fiveS\tabularnewline
$T_{1}$ & \ten & \one & \one\tabularnewline
$T_{2}$ & \one & \ten & \one\tabularnewline
$T_{3}$ & \one & \one & \ten\tabularnewline
\midrule
$\Omega_{12}$ & \24 & \24 & \one\tabularnewline
$\Omega_{13}$ & \24 & \one & \24\tabularnewline
$\Omega_{23}$ & \one & \24 & \24\tabularnewline
\midrule
$H_{1}$ & \five & \one & \one\tabularnewline
$H_{2}$ & \one & \five & \one\tabularnewline
$H_{3}$ & \one & \one & \five\tabularnewline
\bottomrule
\end{tabular}\caption{Minimal content for the general $SU(5)^{3}$ setup. Due to the cyclic symmetry, there are only four independent representations, one for each of the fermions $F_i,T_i$ and the scalars $\Omega_{ij},H_i$. \label{tab:ParticleContent_General}}
\end{table}
\begin{equation}
SU(5)_{1}\times SU(5)_{2}\times SU(5)_{3}\,,\label{eq:SU(5)^3-1}
\end{equation}
where each $SU(5)$ factor is associated to one family of chiral fermions
$i=1,2,3$.
Moreover, we incorporate a cyclic permutation symmetry
$\mathbb{Z}_{3}$ that relates the three $SU(5)$ factors, in the
spirit of the trinification model \cite{Glashow:1984gc}. This implies
that at the high energy GUT scale where $SU(5)^3$ is broken (typically in excess of $10^{16}$ GeV) the gauge couplings
of the three $SU(5)$ factors are equal by cyclic symmetry, such that the
gauge sector is fundamentally described by one gauge coupling. Therefore,
although $SU(5)^{3}$ is a not a simple group, it may be regarded as a unified gauge theory.

The motivation for considering such an  $SU(5)^{3}$ with cyclic symmetry is that it allows gauge non-universal theories of flavour to emerge at low energies\footnote{$SU(5)^{3}$ tri-unification may provide a unified origin for many gauge non-universal theories proposed in the literature to address different questions beyond the flavour puzzle, see e.g.~\cite{Malkawi:1996fs,Shu:2006mm}.} from a gauge universal theory, depending on the symmetry breaking chain. In the first step, $SU(5)^{3}$ may be\footnote{This first step of symmetry breaking is optional, but may be convenient to control the scale of gauge unification as discussed in Section~\ref{sec:gcu}.} broken to three copies of the SM gauge group SM$^3$.
Then at lower energies, SM$^3$ is broken to some universal piece 
$G_{\mathrm{universal}}$ consisting of some diagonal subgroups, together with some remaining family groups 
$G_{1}\times G_{2}\times G_{3}$. If
the Higgs doublet(s) transform non-trivially under the third family
group $G_{3}$, but not under the first nor second, then third family fermions
get natural masses at the electroweak scale, while first and second
family fermions are massless in first approximation. Their small masses
naturally arise from the breaking of the non-universal gauge group
down to the SM, which is the diagonal subgroup, and an approximate
$U(2)^{5}$ flavour symmetry emerges, which is known to provide an efficient suppression of the most dangerous
flavour-violating effects for new physics \cite{Barbieri:2011ci,Allwicher:2023shc}.

At still lower energies, the non-diagonal group factors
$G_{1}\times G_{2}\times G_{3}$
are broken down to their diagonal subgroup, eventually leading to a flavour universal SM gauge group factor. This may happen in stages.
It has been shown that the symmetry
breaking pattern
\begin{equation}
G_{1}\times G_{2}\times G_{3}\rightarrow G_{1+2}\times G_{3}\rightarrow\mathrm{G_{1+2+3}}
\end{equation}
naturally explains the origin of fermion mass hierarchies and the
smallness of quark mixing, while anarchic neutrino mixing may be incorporated
via exotic variations of the type I seesaw mechanism \cite{FernandezNavarro:2023rhv,Fuentes-Martin:2020pww}.

Minimal examples of this class of theories include the tri-hypercharge
model \cite{FernandezNavarro:2023rhv}, which we shall focus on in
this paper, where the universal (diagonal) group consists of the non-Abelian
SM gauge group factors $G_{\mathrm{universal}}=SU(3)_{c}\times SU(2)_{L}$
while the remaining groups are the three gauge weak hypercharge factors
$G_{1}\times G_{2}\times G_{3}=U(1)_{Y_{1}}\times U(1)_{Y_{2}}\times U(1)_{Y_{3}}$.
Another example is the $SU(2)_{L}^{3}$ model \cite{Li:1981nk,Ma:1987ds,Ma:1988dn,Li:1992fi,Muller:1996dj,Chiang:2009kb},
where $G_{\mathrm{universal}}=SU(3)_{c}\times U(1)_{Y}$ and $G_{1}\times G_{2}\times G_{3}=SU(2)_{L1}\times SU(2)_{L2}\times SU(2)_{L3}$.
There also exists the $SU(3)_{c}^{3}$ model \cite{Carone:1995ge}
(which is only able to explain the smallness of quark mixing), where
$G_{\mathrm{universal}}=SU(2)_{L}\times U(1)_{Y}$ and $G_{1}\times G_{2}\times G_{3}=SU(3)_{c1}\times SU(3)_{c2}\times SU(3)_{c3}$.
Variations of these theories have been proposed in recent years, several
of them assuming a possible embedding into (variations of) a Pati-Salam
setup \cite{Bordone:2017bld,Greljo:2018tuh,Allwicher:2020esa,Fuentes-Martin:2020pww,Fuentes-Martin:2022xnb,Davighi:2022bqf,Davighi:2022fer,Davighi:2023iks,Davighi:2023evx}.

All these theories share a common feature: they explain the origin
of the flavour structure of the SM at the price of complicating the
gauge sector, which may now contain up to nine arbitrary gauge couplings. 
We will motivate that $SU(5)^{3}$ as the embedding of general theories
$G_{\mathrm{universal}}\times G_{1}\times G_{2}\times G_{3}$ resolves
this issue, by unifying the complicated gauge sector of these theories
into a single gauge coupling. The main ingredients of our general
setup are as follows:
\begin{itemize}
\item The presence of the $\mathbb{Z}_{3}$ symmetry, which is of fundamental
importance to achieve gauge unification, imposes that the matter content
of the model shall be invariant under cyclic permutations of the three
$SU(5)$ factors. This enforces that each $SU(5)$ factor contains
the same representations of fermions and scalars, i.e.~if the representation
$(\mathbf{A,B,C})$ is included, then $(\mathbf{C,A,B})$ and $(\mathbf{B,C,A})$
must be included too. 
\item Each family of chiral fermions $i$ is embedded in the usual way into
$\mathbf{\overline{5}}$ and $\mathbf{10}$ representations of their
corresponding $SU(5)_{i}$ factor,
that we denote as $F_{i}=(d_{i}^{c},\ell_{i})\sim \mathbf{\overline{5}}_{i}$ and $T_{i}=(q_{i},u_{i}^{c},e_{i}^{c})\sim \mathbf{10}_{i}$ as shown in Table~\ref{tab:ParticleContent_General}.
This choice is naturally consistent with the $\mathbb{Z}_{3}$ symmetry.
\item In a similar manner, three Higgs doublets $H_{1}$, $H_{2}$ and $H_{3}$
are embedded into $\mathbf{5}$ representations, one
for each $SU(5)_{i}$ factor. Notice that in non-universal theories
of flavour it is commonly assumed the existence of only one Higgs
doublet $H_{3}$, which transforms only under the third site in order
to explain the heaviness of the third family. This way, the $SU(5)^{3}$
framework involves the restriction of having three Higgses rather
than only one, but we will argue that if the $\mathbb{Z}_{3}$ symmetry
is broken below the GUT scale, then only the third family Higgs $H_{3}$
may be light and perform electroweak symmetry breaking, while $H_{1}$
and $H_{2}$ are heavier and may play the role of heavy messengers
for the effective Yukawa couplings of the light families.
\item Higgs scalars in bi-representations connecting the different sites
may be needed to generate the SM flavour structure at the level of
the $G_{\mathrm{universal}}\times G_{1}\times G_{2}\times G_{3}$
theory, e.g.~$(\mathbf{2,\overline{2}})$ scalars in $SU(2)_{L}^{3}$
or $(Y,-Y)$ scalars in tri-hypercharge (the so-called hyperons). These can be embedded in
the associated bi-representations of $SU(5)^{3}$, e.g.~$(\mathbf{5,\overline{5}})$
scalars, $(\mathbf{10,\overline{10}})$ scalars and so on. In Appendix~\ref{sec:Hyperons}
we tabulate all such scalars from $SU(5)^{3}$ representations with dimension up to $\mathbf{45}$, along with
the hyperons that they generate at low energies.

\item Finally, three scalar fields in bi-adjoint representations of each $SU(5)$,
$\Omega_{ij}$, spontaneously break the tri-unification symmetry. The three 
$\Omega_{ij}$ are enough to perform \textit{both} horizontal and vertical breaking of the three $SU(5)$ groups at the GUT scale, down to 
the non-universal gauge group $G_{\mathrm{universal}}\times G_{1}\times G_{2}\times G_{3}$ of choice
that later explains the flavour structure of the SM (e.g.~tri-hypercharge
or $SU(2)_{L}^{3}$). Another possibility that we will explore is
breaking $SU(5)^{3}$ first to three copies of the SM (one for each
family) and then to $G_{\mathrm{universal}}\times G_{1}\times G_{2}\times G_{3}$
in a second step. 
\end{itemize}

To summarise, the general pattern of symmetry breaking we assume is as follows\footnote{One should note that none of the individual groups, $SU(3)$, $SU(2)$ or $U(1)$, in each SM$_i$ group correspond to the SM's $SU(3)_c$, $SU(2)_L$ or $U(1)_Y$. The latter emerge after symmetry breaking from the diagonal sub-groups of the former. Nevertheless, we will denote each $\left(SU(3) \times SU(2) \times U(1)\right)_i$ as SM$_i$ and the total $\left(SU(3) \times SU(2) \times U(1)\right)^3$ group as SM$^3$ for the sake of brevity.},
\begin{flalign}
SU(5)^{3} & \rightarrow\mathrm{SM_{1}}\times\mathrm{SM_{2}}\times\mathrm{SM_{3}}\\
 & \rightarrow G_{\mathrm{universal}}\times G_{1}\times G_{2}\times G_{3}\\
 & \rightarrow G_{\mathrm{universal}}\times G_{1+2}\times G_{3}\\
 & \rightarrow\mathrm{SM_{1+2+3}}\,,
\end{flalign}
where the $\mathrm{SM}^{3}$ step is optional but may be convenient
to achieve unification. In particular, the first step of symmetry breaking makes use of three SM singlets contained in $\Omega_{ij}$, while the second step may be performed via the remaining degrees of freedom in $\Omega_{ij}$, depending on the details of the low energy gauge theory that survives. The two final breaking steps are performed by Higgs scalars connecting the different sites that need to be specified for each particular model.

Beyond the general considerations listed in this section, when building a specific
model one needs to choose the symmetry group $G_{\mathrm{universal}}\times G_{1}\times G_{2}\times G_{3}$,
and add explicit scalars and/or fermion messengers that mediate the effective Yukawa
couplings of light fermions.

Finally, one needs to study the Renormalization Group Equations (RGEs) of the various gauge couplings at the different steps all the way up to the $SU(5)^{3}$ scale where all gauge couplings need to unify. This is not a simple task, but we shall see that the relatively light messengers required to generate the effective
Yukawa couplings, along with the presence of the approximate $\mathbb{Z}_{3}$
symmetry at low energies, may naturally help to achieve unification.
In the following, we shall illustrate this by describing a working
example of the $SU(5)^{3}$ framework based on tri-hypercharge \cite{FernandezNavarro:2023rhv},
where the various gauge couplings of the tri-hypercharge model unify
at the GUT scale into a single gauge coupling.

\begin{table}[H]
\centering{}%
\vspace{-1.5cm}
\begin{tabular}{cccc}
\toprule 
\textbf{Field}  & $\boldsymbol{SU(5)_{1}}$  & $\boldsymbol{SU(5)_{2}}$  & $\boldsymbol{SU(5)_{3}}$\tabularnewline
\midrule 
$F_{1}$  & \fiveS  & \one  & \one\tabularnewline
$F_{2}$  & \one  & \fiveS  & \one\tabularnewline
$F_{3}$  & \one  & \one  & \fiveS\tabularnewline
$T_{1}$  & \ten  & \one  & \one\tabularnewline
$T_{2}$  & \one  & \ten  & \one\tabularnewline
$T_{3}$  & \one  & \one  & \ten\tabularnewline
\midrule 
\rowcolor{yellow!10} $\chi_{1}$  & \ten  & \one  & \one\tabularnewline
\rowcolor{yellow!10} $\chi_{2}$  & \one  & \ten  & \one\tabularnewline
\rowcolor{yellow!10} $\chi_{3}$  & \one  & \one  & \ten\tabularnewline
\midrule
\rowcolor{yellow!10} $\Xi_{0}$ & \one & \one & \one\tabularnewline
\rowcolor{yellow!10} $\Xi_{12}$ & \five  & \fiveS  & \one\tabularnewline
\rowcolor{yellow!10} $\Xi_{13}$ & \fiveS  & \one  & \five\tabularnewline
\rowcolor{yellow!10} $\Xi_{23}$ & \one  & \five  & \fiveS\tabularnewline
\midrule
\rowcolor{yellow!10} $\Sigma_{\mathrm{atm}}$ & \one  & \ten & \tenS \tabularnewline
\rowcolor{yellow!10} $\Sigma_{\mathrm{sol}}$ & \ten & \one  & \tenS \tabularnewline
\rowcolor{yellow!10} $\Sigma_{\mathrm{cyclic}}$ & \ten & \tenS  & \one \tabularnewline
\midrule 
$\Omega_{12}$ & \24 & \24 & \one\tabularnewline
$\Omega_{13}$ & \24 & \one & \24\tabularnewline
$\Omega_{23}$ & \one & \24 & \24\tabularnewline
\midrule 
$H_{1}^{u}$  & \five  & \one  & \one\tabularnewline
$H_{2}^{u}$  & \one  & \five  & \one\tabularnewline
$H_{3}^{u}$  & \one  & \one  & \five\tabularnewline
$H_{1}^{\mathbf{\overline{5}}}$  & \fiveS  & \one  & \one\tabularnewline
$H_{2}^{\mathbf{\overline{5}}}$  & \one  & \fiveS  & \one\tabularnewline
$H_{3}^{\mathbf{\overline{5}}}$  & \one  & \one  & \fiveS\tabularnewline
$H_{1}^{\mathbf{45}}$  & $\mathbf{45}$  & \one  & \one\tabularnewline
$H_{2}^{\mathbf{45}}$  & \one  & $\mathbf{45}$  & \one\tabularnewline
$H_{3}^{\mathbf{45}}$  & \one  & \one  & $\mathbf{45}$\tabularnewline
\midrule 
$\Phi_{12}^{F}$  & \five  & \fiveS  & \one\tabularnewline
$\Phi_{13}^{F}$  & \fiveS  & \one  & \five\tabularnewline
$\Phi_{23}^{F}$  & \one  & \five  & \fiveS\tabularnewline
$\Phi_{12}^{T}$  & \tenS  & \ten  & \one\tabularnewline
$\Phi_{13}^{T}$  & \ten  & \one  & \tenS\tabularnewline
$\Phi_{23}^{T}$  & \one  & \tenS  & \ten\tabularnewline
\midrule
$\Phi_{12}^{\mathbf{45}}$  & \one  & $\mathbf{\overline{45}}$  & $\mathbf{45}$ \tabularnewline
$\Phi_{13}^{\mathbf{45}}$  & $\mathbf{\overline{45}}$  & \one  & $\mathbf{45}$ \tabularnewline
$\Phi_{12}^{\mathbf{45}}$  & $\mathbf{\overline{45}}$  & $\mathbf{45}$  & \one \tabularnewline
$\Phi^{TFT}$  & \ten  & \five  & \ten \tabularnewline
$\Phi^{FTT}$  & \five  & \ten  & \ten \tabularnewline
$\Phi^{TTF}$  & \ten  & \ten  & \five \tabularnewline
\bottomrule
\end{tabular}\caption{Fermion and scalar particle content and representations under $SU(5)^{3}$.
$F_{i}$ and $T_{i}$ include the chiral fermions of the SM in the
usual way, while $\chi_{i}$, $\xi$'s and $\Xi$'s (highlighted in
yellow) are vector-like fermions, thus the conjugate partners must
be considered. $\Omega$'s, $H$'s and $\Phi$'s are scalars. 
\label{tab:ParticleContent_model}}
\end{table}

\section{An example \texorpdfstring{$\boldsymbol{SU(5)^{3}}$}{SU(5)^{3}} unification model breaking to tri-hypercharge\label{sec:The-Tri-Hypercharge-model}}

We now turn to the main example of interest, namely $G_{\mathrm{universal}}=SU(3)_{c}\times SU(2)_{L}$ (the diagonal non-Abelian SM gauge group factors) 
with $G_{1}\times G_{2}\times G_{3}=U(1)_{Y_{1}}\times U(1)_{Y_{2}}\times U(1)_{Y_{3}}$ 
(the tri-hypercharge model).
In this example, the basic idea is that $SU(5)^{3}$ breaks, via a sequence of scales, to 
the low energy (well below the GUT scale) tri-hypercharge gauge group with a separate gauged weak
hypercharge for each fermion family, 
\begin{equation}
SU(5)^{3} \rightarrow \dots \rightarrow
SU(3)_{c}\times SU(2)_{L}\times U(1)_{Y_{1}}\times U(1)_{Y_{2}}\times U(1)_{Y_{3}}\,.
\label{GUTbreaking}
\end{equation}
In \cite{FernandezNavarro:2023rhv} it was shown that the low energy tri-hypercharge model can
naturally generate the flavour structure of the SM if spontaneously
broken to SM hypercharge
in a convenient way. The minimal setup involves the vacuum expectation values (VEVs) of
the new Higgs ``hyperons'' 
\begin{equation}
\phi_{q12}\sim(\mathbf{1,1})_{(-1/6,1/6,0)},\qquad\phi_{q23}\sim(\mathbf{1,1})_{(0,-1/6,1/6)},\qquad\phi_{\ell23}\sim(\mathbf{1,1})_{(0,1/2,-1/2)}\,.
\end{equation}
At the GUT scale, the hyperons are embedded into
bi-$\mathbf{\overline{5}}$ and bi-$\mathbf{10}$ representations
of $SU(5)^{3}$ expressed as $\Phi_{ij}^{T,F}$, which must preserve
the cyclic symmetry, as shown in Table~\ref{tab:ParticleContent_model}.
Although this involves the appearance of many hyperons (and other
scalars) beyond the minimal set of hyperons that we need, we shall
assume that only the desired hyperons get a VEV (and the rest of scalars
may remain very heavy). Moreover, the $SU(5)^{3}$ framework also
poses constraints on the possible family hypercharges of the hyperons, as collected in Appendix~\ref{sec:Hyperons}.
For the $SU(5)^{3}$ setup, it is convenient to add 
\begin{equation}
\phi_{q13}\sim(\mathbf{1,1})_{(-1/6,0,1/6)}\,,\qquad\phi_{\ell13}\sim(\mathbf{1,1})_{(1/2,0,-1/2)}\,,
\end{equation}
which are anyway required by the cyclic symmetry, to the set of hyperons
which get a VEV.

The hyperons allow to write a set of non-renormalisable operators
that provide effective Yukawa couplings for light fermions, as described
in~\cite{FernandezNavarro:2023rhv} by working in an effective field theory (EFT) framework.
However, in our unified model, we need to introduce heavy messengers
that mediate such effective operators in order to obtain a UV complete
setup. For this, we add one set of vector-like fermions transforming
in the $\mathbf{10}$ representation for each $SU(5)$ factor, i.e.~$\chi_{i}\sim\mathbf{10}_{i}$
and $\overline{\chi}_{i}\sim\overline{\mathbf{10}}_{i}$. We shall
assume that only the quark doublets $Q_{i}\sim(\mathbf{3,2})_{1/6_{i}}$
and $\overline{Q}_{i}\sim(\mathbf{\overline{3},2})_{-1/6_{i}}$ are
relatively light and play a role in the effective Yukawa couplings,
while the remaining degrees of freedom in $\chi_{i}$ and $\overline{\chi}_{i}$
remain very heavy,
\begin{equation}
\chi_{i}\rightarrow Q_{i}\sim(\mathbf{3,2})_{1/6_{i}}\,,\qquad\overline{\chi}_{i}\rightarrow\overline{Q}_{i}\sim(\mathbf{\overline{3},2})_{-1/6_{i}}\,.
\end{equation} We shall see that $Q_{i}$ and $\overline{Q}_{i}$
also contribute to the RGEs in the desired way to achieve gauge unification.
The full field content of this model also includes extra vector-like fermions $\Sigma$ and $\Xi$ as shown in Table~\ref{tab:ParticleContent_model}. These play a role in the origin of neutrino masses as discussed in Section~\ref{subsec:Neutrinos}.

Finally, beyond the minimal set of Higgs doublets introduced in Section~\ref{sec:General-framework},
we shall introduce here three pairs of $\mathbf{5}$, $\overline{\mathbf{5}}$
and $\mathbf{45}$ Higgs representations preserving the cyclic symmetry.
The doublets in the $\overline{\mathbf{5}}$ and $\mathbf{45}$ mix,
leaving light linear combinations that couple differently to down-quarks
and charged leptons in the usual way \cite{Georgi:1979df}, which we denote as $H_{i}^{d}$.

Therefore, below the GUT scale we effectively have three pairs of
Higgs doublets $H_{1}^{u,d}$, $H_{2}^{u,d}$ and $H_{3}^{u,d}$,
such that the $u$- and $d$- labeled Higgs only couple to up-quarks
(and neutrinos) and to down-quarks and charged leptons, respectively,
in the spirit of the type II two Higgs doublet model. This choice
is motivated to explain the mass hierarchies between the different
charged sectors, as originally identified in \cite{FernandezNavarro:2023rhv},
and could be enforced e.g.~by a $\mathbb{Z}_{2}$ discrete symmetry.
We assume that the third family Higgs $H_{3}^{u,d}$ are the lightest,
they perform electroweak symmetry breaking and provide Yukawa couplings
for the third family with $\mathcal{O}(1)$ coefficients if $\tan\beta\approx20$.
In contrast, we assume that the Higgs $H_{1}^{u,d}$, $H_{2}^{u,d}$
have masses above the TeV (but much below the GUT scale) and act as
messengers of the effective Yukawa couplings for the light families.

\begin{figure}
\includegraphics[scale=0.92]{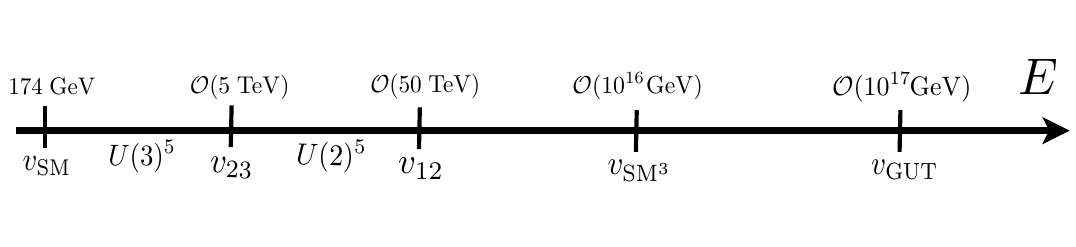}
\caption{Diagram showing the different scales of spontaneous symmetry breaking in our example model (see also Eqs.(\ref{eq:GUT_breaking}-\ref{eq:SM})),
along with the accidental, approximate flavour symmetries ($U(3)^{5}$ and $U(2)^{5}$) that arise at low energies.}
\label{fig:scales}

\end{figure}

In detail, we assume that the $SU(5)^{3}$ group is broken
down to the SM through the following symmetry breaking chain
\begin{flalign}
SU(5)^{3} & \xrightarrow{v_{\mathrm{GUT}}} 
\mathrm{SM_{1}}\times\mathrm{SM_{2}}\times\mathrm{SM_{3}}\label{eq:GUT_breaking}\\
 & \xrightarrow{v_{\rm SM^3}} 
 SU(3)_{1+2+3}\times SU(2)_{1+2+3}\times U(1)_{1}\times U(1)_{2}\times U(1)_{3}\\
 & \xrightarrow{v_{12}} 
 SU(3)_{1+2+3}\times SU(2)_{1+2+3}\times U(1)_{1+2}\times U(1)_{3} \label{eq:23_breaking}\\
 & \xrightarrow{v_{23}} 
 SU(3)_{1+2+3}\times SU(2)_{1+2+3}\times U(1)_{1+2+3}\label{eq:SM}\,.
\end{flalign}
The $SU(5)^{3}$ breaking happens at the GUT scale, 
while the tri-hypercharge
breaking may happen as low as the TeV scale, as allowed by current
data \cite{FernandezNavarro:2023rhv}, while the $\mathrm{SM}^{3}$ breaking step is optional but may be convenient
to achieve unification, and may be regarded as free parameter. This
second breaking step is performed by the $SU(3)_{i}$ octets and $SU(2)_{i}$
triplets contained in $\Omega_{ij}\sim\mathbf{24}_{i}$. See also Fig.~\ref{fig:scales}
for an illustrative diagram.

We shall show that within this setup, achieving gauge unification
just requires further assuming that three colour octets that live
in $\Omega_{ij}$ are light, while the remaining
degrees of freedom of the bi-adjoints remain very heavy. Before that,
we shall study in detail how our model explains the origin of the
flavour structure of the SM.

\subsection{Charged fermion mass hierarchies and quark mixing \label{subsec:Charged_fermions}}
The Higgs doublets in the cyclic $\mathbf{\overline{5}}$ and $\mathbf{45}$
split the couplings of down-quarks and charged leptons in the usual
way \cite{Georgi:1979df}. We denote as $H_{i}^{d}$ the linear combinations that
remain light, with their effective couplings to down-quarks and charged
leptons given by
\begin{equation}
\widetilde{y}_{i}^{d}H_{i}^{d}T_{i}F_{i}\rightarrow y_{i}^{d}H_{i}^{d}q_{i}d_{i}^{c}+y_{i}^{e}H_{i}^{d}\ell_{i}e_{i}^{c},
\end{equation}
where
\begin{equation}
y_{i}^{d}=y_{i}^{\mathbf{\overline{5}}}+y_{i}^{\mathbf{\overline{45}}}\,,\qquad y_{i}^{e}=y_{i}^{\mathbf{\overline{5}}}-3y_{i}^{\mathbf{\overline{45}}}\,.
\end{equation}
We focus now on the following set of couplings involving the hyperons,
the vector-like fermions $\chi_{i}$ and the light linear combinations
of Higgs doublets,
\begin{align}
\mathcal{L} &\supset x_{ij} \, \Phi_{ij}^{T}T_{i}\overline{\chi}_{j} + z_{i}^{u} \, H_{i}^{u}\chi_{i}T_{i} + z_{i}^{d} \, H_{i}^{d}\chi_{i}F_{i} \nonumber \\
&+y_{i}^{u} \, H_{i}^{u}T_{i}T_{i} + \widetilde{y}_{i}^{d} \, H_{i}^{d}T_{i}F_{i} + f_{ij}^{u} \, H_{i}^{u}\widetilde{H}_{j}^{u}\widetilde{\Phi}_{ij}^{F} + f_{ij}^{d} \, H_{i}^{d}\widetilde{H}_{j}^{d}\Phi_{ij}^{F} \, ,
\end{align}
where $i,j=1,2,3$, $f_{ij}^{u,d}$ have mass dimension and the rest of the couplings
are dimensionless. After integrating out the heavy vector-like fermions
$\chi_{i}$, $\overline{\chi}_{i}$ and Higgs doublets $H_{1,2}^{u,d}$,
we obtain the following set of effective Yukawa couplings,
\begin{flalign}
\mathcal{L} & =\begin{pmatrix}q_{1} & q_{2} & q_{3}\end{pmatrix}\begin{pmatrix}{\displaystyle c_{11}^{u}\frac{{\phi}_{\ell13}}{M_{H_{1}^{u}}}} & {\displaystyle c_{12}^{u}\frac{{\phi}_{\ell23}}{M_{H_{2}^{u}}}\frac{{\phi}_{q12}}{M_{Q_{2}}}} & {\displaystyle c_{13}^{u}\frac{{\phi}_{q13}}{M_{Q_{3}}}}\\
\vspace{-0.1cm} & \vspace{-0.1cm} & \vspace{-0.1cm}\\
{\displaystyle c_{21}^{u}\frac{{\phi}_{\ell13}}{M_{H_{1}^{u}}}\frac{{\phi}_{q12}}{M_{Q_{1}}}} & {\displaystyle c_{22}^{u}\frac{{\phi}_{\ell23}}{M_{H_{2}^{u}}}} & {\displaystyle c_{23}^{u}\frac{{\phi}_{q23}}{M_{Q_{3}}}}\\
\vspace{-0.1cm} & \vspace{-0.1cm} & \vspace{-0.1cm}\\
{\displaystyle c_{31}^{u}\frac{{\phi}_{\ell13}}{M_{H_{1}^{u}}}\frac{{\widetilde{\phi}}_{q13}}{M_{Q_{1}}}} & {\displaystyle c_{32}^{u}\frac{{\phi}_{\ell23}}{M_{H_{2}^{u}}}\frac{{\widetilde{\phi}}_{q23}}{M_{Q_{2}}}} & {\displaystyle c_{33}^{u}}
\end{pmatrix}\begin{pmatrix}u_{1}^{c}\\
u_{2}^{c}\\
u_{3}^{c}
\end{pmatrix}H_{3}^{u}\label{eq:eff_Yukawa_ups}\\
 & +\begin{pmatrix}q_{1} & q_{2} & q_{3}\end{pmatrix}\begin{pmatrix}{\displaystyle c_{11}^{d}\frac{{\widetilde{\phi}}_{\ell13}}{M_{H_{1}^{d}}}} & {\displaystyle c_{12}^{d}\frac{{\widetilde{\phi}}_{\ell23}}{M_{H_{2}^{d}}}\frac{{\phi}_{q12}}{M_{Q_{2}}}} & {\displaystyle c_{13}^{d}\frac{{\phi}_{q13}}{M_{Q_{3}}}}\\
\vspace{-0.1cm} & \vspace{-0.1cm} & \vspace{-0.1cm}\\
{\displaystyle c_{21}^{d}\frac{{\widetilde{\phi}}_{\ell13}}{M_{H_{1}^{d}}}\frac{{\phi}_{q12}}{M_{Q_{1}}}} & {\displaystyle c_{22}^{d}\frac{{\widetilde{\phi}}_{\ell23}}{M_{H_{2}^{d}}}} & {\displaystyle c_{23}^{d}\frac{{\phi}_{q23}}{M_{Q_{3}}}}\\
\vspace{-0.1cm} & \vspace{-0.1cm} & \vspace{-0.1cm}\\
{\displaystyle c_{31}^{d}\frac{{\widetilde{\phi}}_{\ell13}}{M_{H_{1}^{d}}}\frac{{\phi}_{q13}}{M_{Q_{1}}}} & {\displaystyle c_{32}^{d}\frac{{\widetilde{\phi}}_{\ell23}}{M_{H_{2}^{d}}}\frac{{\phi}_{q23}}{M_{Q_{2}}}} & {\displaystyle c_{33}^{d}}
\end{pmatrix}\begin{pmatrix}d_{1}^{c}\\
d_{2}^{c}\\
d_{3}^{c}
\end{pmatrix}H_{3}^{d}\\
 & +\begin{pmatrix}\ell_{1} & \ell_{2} & \ell_{3}\end{pmatrix}\begin{pmatrix}{\displaystyle c_{11}^{e}\frac{{\widetilde{\phi}}_{\ell13}}{M_{H_{1}^{d}}}} & {\displaystyle 0} & {\displaystyle 0}\\
{\displaystyle 0} & {\displaystyle c_{22}^{e}\frac{{\widetilde{\phi}}_{\ell23}}{M_{H_{2}^{d}}}} & {\displaystyle 0}\\
{\displaystyle 0} & {\displaystyle 0} & {\displaystyle c_{33}^{e}}
\end{pmatrix}\begin{pmatrix}e_{1}^{c}\\
e_{2}^{c}\\
e_{3}^{c}
\end{pmatrix}H_{3}^{d}+\mathrm{h.c.}\,,\label{eq:Eff_Yukawa_charged_leptons}
\end{flalign}
where the dimensionless coefficients $c_{ij}^{u,d,e}$ are given by
\begin{equation}
c_{ij}^{e}=\mathrm{diag}\left({\displaystyle y_{1}^{e}\frac{f_{13}^{d}}{M_{H_{1}}^{d}}},{\displaystyle y_{2}^{e}\frac{f_{23}^{d}}{M_{H_{2}}^{d}}},{\displaystyle y_{3}^{e}}\right)\,.
\end{equation}
\begin{equation}
c_{ij}^{u,d}={\displaystyle \begin{pmatrix}{\displaystyle y_{1}^{u,d}\frac{f_{13}^{u,d}}{M_{H_{1}}^{u,d}}} & {\displaystyle x_{12}y_{2}^{u,d}\frac{f_{23}^{u,d}}{M_{H_{2}}^{u,d}}} & {\displaystyle x_{13}z_{3}^{u,d}}\\
\vspace{-0.1cm} & \vspace{-0.1cm} & \vspace{-0.1cm}\\
{\displaystyle x_{21}y_{1}^{u,d}\frac{f_{13}^{u,d}}{M_{H_{1}}^{u,d}}} & {\displaystyle y_{2}^{u,d}\frac{f_{23}^{u,d}}{M_{H_{2}}^{u,d}}} & {\displaystyle x_{23}z_{3}^{u,d}}\\
\vspace{-0.1cm} & \vspace{-0.1cm} & \vspace{-0.1cm}\\
{\displaystyle x_{31}y_{1}^{u,d}\frac{f_{13}^{u,d}}{M_{H_{1}}^{u,d}}} & {\displaystyle x_{32}y_{2}^{u,d}\frac{f_{23}^{u,d}}{M_{H_{2}}^{u,d}}} & {\displaystyle y_{3}^{u,d}}
\end{pmatrix}}\,.
\end{equation}

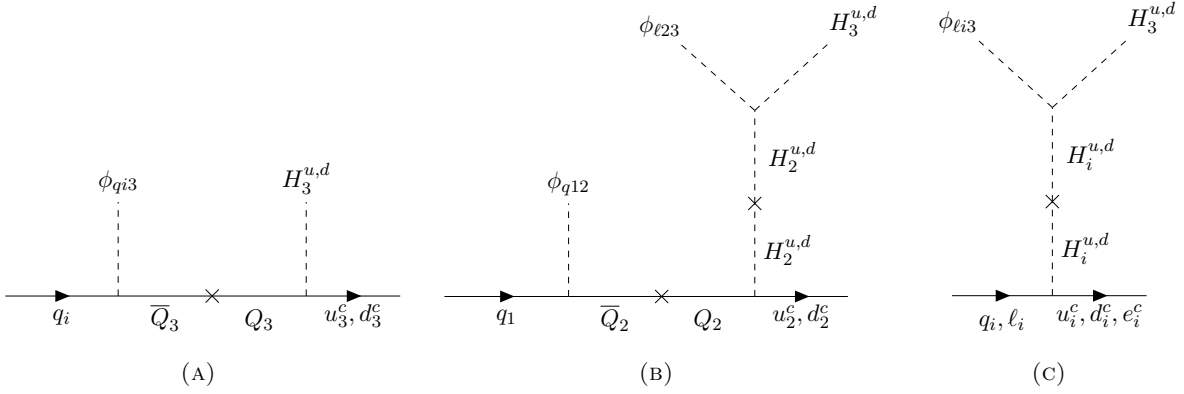
\begin{figure}[t]
\subfloat[]{\resizebox{.33\textwidth}{!}{
\begin{tikzpicture}
	\begin{feynman}
		\vertex (a);
		\vertex [right=18mm of a] (b);
		\vertex [right=of b] (c);
		\vertex [right=of c] (d);
		\vertex [right=of d] (e);
		\vertex [above=of b] (f1) {\(\phi_{qi3}\)};
		\vertex [above=of d] (f2) {\(H_{3}^{u,d}\)};
		\diagram* {
			(a) -- [fermion, edge label'=\(q_{i}\), inner sep=6pt] (b) -- [scalar] (f1),
			(b) -- [edge label'=\(\overline{Q}_{3}\)] (c),
			(c) -- [edge label'=\(Q_{3}\),inner sep=6pt, insertion=0] (d) -- [scalar] (f2),
			(d) -- [fermion, edge label'=\(u_{3}^{c}{,}\,d_{3}^{c}\)] (e),
	};
	\end{feynman}
\end{tikzpicture}
}

}$\;$\subfloat[]{\begin{centering}
\resizebox{.36\textwidth}{!}{ \begin{tikzpicture}
	\begin{feynman}
		\vertex (a);
		\vertex [right=20mm of a] (b);
        \vertex [above=of b] (h1) {\(\phi_{q12}\)};
		\vertex [right=of b] (c);
        \vertex [above=of c] (h2);
		\vertex [right=of c] (d);
        \vertex [right=of d] (e);
		\vertex [above=of d] (f1);
		\vertex [above=of f1] (f2);
		\vertex [above left=of f2] (g1) {\(\phi_{\ell 23}\)};
		\vertex [above right=of f2] (g2) {\(H^{u,d}_{3}\)};
		\diagram* {
			(a) -- [fermion, edge label'=\(q_{1}\), inner sep=6pt] (b) -- [scalar] (h1),
			(b) -- [edge label'=\(\overline{Q}_{2}\)] (c) -- [edge label'=\(Q_{2}\), inner sep=6pt, insertion=0] (d),
            (d) -- [fermion, edge label'=\(u_{2}^{c}{,}\,d_{2}^{c}\)] (e),
			(d) -- [scalar, edge label'=\(H_{2}^{u,d}\)] (f1) -- [scalar, edge label'=\(H_{2}^{u,d}\), inner sep=6pt, insertion=0] (f2),
			(f2) -- [scalar] (g1),
			(f2) -- [scalar] (g2),
	};
	\end{feynman}
\end{tikzpicture}}
\par\end{centering}
}$\;$\subfloat[]{\begin{centering}
\resizebox{.21\textwidth}{!}{ \begin{tikzpicture}
	\begin{feynman}
		\vertex (a);
		\vertex [right=16mm of a] (b);
		\vertex [right=of b] (c);
		\vertex [above=of b] (f1);
		\vertex [above=of f1] (f2);
		\vertex [above left=of f2] (g1) {\(\phi_{\ell i3}\)};
		\vertex [above right=of f2] (g2) {\(H^{u,d}_{3}\)};
		\diagram* {
			(a) -- [fermion, edge label'=\(q_{i}{,}\,\ell_{i}\), inner sep=6pt] (b) -- [fermion, edge label'=\(u^{c}_{i}{,}\,d^{c}_{i}{,}\,e^{c}_{i}\)] (c),
			(b) -- [scalar, edge label'=\(H_{i}^{u,d}\)] (f1),
			(f1) -- [scalar, edge label'=\(H_{i}^{u,d}\),inner sep=6pt, insertion=0] (f2),
			(f2) -- [scalar] (g1),
			(f2) -- [scalar] (g2),
	};
	\end{feynman}
\end{tikzpicture}}
\par\end{centering}
}\caption[]{Diagrams in the model which lead to the origin of light charged fermion masses
and quark mixing, where $i=1,2$. \label{fig:charged_fermions_diagrams}}
\end{figure}

It is clear that third family charged fermions get
their masses from $\mathcal{O}(1)$ Yukawa couplings to the Higgs
doublets $H_{3}^{u,d}$, where the mass hierarchies $m_{b,\tau}/m_{t}$
are explained via $\tan\beta\approx \lambda^{-2}$, where $\lambda\simeq 0.224$ is the
Wolfenstein parameter. In contrast, quark mixing
and the masses of first and second family charged fermions arise from
effective Yukawa couplings involving the heavy messengers of the model,
once the hyperons develop their VEVs. The heavy Higgs doublets $H_{1}^{u,d}$
and $H_{2}^{u,d}$ play a role in the origin of the family mass hierarchies,
while the origin of quark mixing involves both the heavy Higgs and
the vector-like quarks $Q_{i}$ and $\overline{Q}_{i}$, as shown
in Fig.~\ref{fig:charged_fermions_diagrams}. We fix the various
$\left\langle \phi\right\rangle /M$ ratios in terms of the Wolfenstein
parameter $\lambda\simeq0.224$
\begin{equation}
\frac{\left\langle \phi_{q23}\right\rangle }{M_{Q_{i}}}\approx\lambda^{2}\,,\qquad\frac{\left\langle \phi_{q13}\right\rangle }{M_{Q_{i}}}\approx\lambda^{3},\qquad\frac{\left\langle \phi_{\ell23}\right\rangle }{M_{H_{2}^{u,d}}}\approx\lambda^{3}\,,\qquad\frac{\left\langle \phi_{q12}\right\rangle }{M_{Q_{i}}}\approx\lambda\,,\qquad\frac{\left\langle \phi_{\ell23}\right\rangle }{M_{H_{1}^{u,d}}}\approx\lambda^{6}\,.\label{eq:VEV_M_ratios}
\end{equation}
We notice that the tiny masses of the first family are explained via
the hierarchies of Higgs messengers 
\begin{equation}
M_{H_{3}^{u,d}}\ll M_{H_{2}^{u,d}}\ll M_{H_{1}^{u,d}}\,,
\end{equation}
in the spirit of messenger dominance \cite{Ferretti:2006df}. In other
words, the heavy Higgs doublets $H_{1}^{u,d}$ and $H_{2}^{u,d}$
can be thought of gaining small effective VEVs from mixing with $H_{3}^{u,d}$,
which are light and perform electroweak symmetry breaking, and these
effective VEVs provide naturally small masses for light charged fermions.
This is in contrast with the original spirit of tri-hypercharge, where
the $m_{1}/m_{2}$ mass hierarchies find their natural origin due
to the higher dimension of the effective Yukawa couplings involving
the first family \cite{FernandezNavarro:2023rhv}. However, we note
that in the $SU(5)^{3}$ framework, the three pairs of Higgs doublets
$H_{i}^{u,d}$ are required by the $\mathbb{Z}_{3}$ symmetry, hence
it seems natural that they play a role on the origin of fermion masses.
Moreover, the introduction of these Higgs provides a very minimal
framework to UV-complete the effective Yukawa couplings of tri-hypercharge,
which otherwise would require a much larger amount of heavy messengers
that are not desired, as they may
enhance too much the RGE of the gauge couplings, eventually leading to a non-perturbative
gauge coupling at the GUT scale.

The numerical values for the ratios in Eq.~(\ref{eq:VEV_M_ratios})
provide the following Yukawa textures (ignoring dimensionless coefficients)
\begin{flalign}
\mathcal{L} & =\begin{pmatrix}q_{1} & q_{2} & q_{3}\end{pmatrix}\begin{pmatrix}\lambda^{6} & \lambda^{4} & \lambda^{3}\\
\lambda^{7} & \lambda^{3} & \lambda^{2}\\
\lambda^{9} & \lambda^{5} & 1
\end{pmatrix}\begin{pmatrix}u_{1}^{c}\\
u_{2}^{c}\\
u_{3}^{c}
\end{pmatrix}v_{\mathrm{SM}}\label{eq:eff_Yukawa_ups-1}\\
 & +\begin{pmatrix}q_{1} & q_{2} & q_{3}\end{pmatrix}\begin{pmatrix}\lambda^{6} & \lambda^{4} & \lambda^{3}\\
\lambda^{7} & \lambda^{3} & \lambda^{2}\\
\lambda^{9} & \lambda^{5} & 1
\end{pmatrix}\begin{pmatrix}d_{1}^{c}\\
d_{2}^{c}\\
d_{3}^{c}
\end{pmatrix}\lambda^{2} \, v_{\mathrm{SM}}\\
 & +\begin{pmatrix}\ell_{1} & \ell_{2} & \ell_{3}\end{pmatrix}\begin{pmatrix}\lambda^{6} & 0 & 0\\
0 & \lambda^{3} & 0\\
0 & 0 & 1
\end{pmatrix}\begin{pmatrix}e_{1}^{c}\\
e_{2}^{c}\\
e_{3}^{c}
\end{pmatrix}\lambda^{2} \, v_{\mathrm{SM}}+\mathrm{h.c.}\,,\label{eq:Eff_Yukawa_charged_leptons-1}
\end{flalign}
where $v_{\mathrm{SM}}$ is the usual SM electroweak VEV and the fit of the up-quark mass may be improved by assuming
a mild difference between $M_{H_{1}^{u}}$ and $M_{H_{1}^{d}}$.
In general, the alignment of the CKM matrix is not predicted but depends on the choice of dimensionless
coefficients and on the difference between $M_{H_{2}^{u}}$ and $M_{H_{2}^{d}}$.
Any charged lepton mixing is suppressed by the very heavy masses of the required
messengers contained in $\chi_{i}$ and
$\overline{\chi}_{i}$, leading to the off-diagonal zeros in Eq.~\eqref{eq:Eff_Yukawa_charged_leptons-1}, in such a way that 
the PMNS matrix must dominantly arise from the neutrino sector, as we shall see.
We notice that a mild hierarchy of dimensionless couplings $y^e_{1}/y^d_{1}\approx \lambda^{1.4}$
may be needed to account for the mass hierarchy between the down-quark and the electron.

The larger suppression of the (2,1), (3,1) and (3,2) entries in the quark Yukawa textures ensures
a significant suppression of right-handed quark mixing. This is
a very desirable feature, given the strong phenomenological constraints
on right-handed flavour-changing currents \cite{UTfit:2007eik,Isidori:2014rba}.
This way, we expect the model to reproduce the low energy phenomenology
of Model 2 in \cite{FernandezNavarro:2023rhv}, where the VEVs of
the 23 and 13 hyperons may be as low as the TeV scale, while the VEVs
of the 12 hyperons may be as low as 50 TeV or so. In this manner,
we provide the following benchmark values for the mass scales involved
in the flavour sector\footnote{We note that all VEVs and masses listed here may vary by $\mathcal{O}(1)$
factors, as naturally expected, without affecting our final conclusions.}
\begin{equation}
\left\langle \phi_{q23}\right\rangle \approx\left\langle \phi_{q13}\right\rangle \approx\left\langle \phi_{\ell23}\right\rangle \approx\left\langle \phi_{\ell13}\right\rangle \sim\mathcal{O}(5\,\mathrm{TeV})\,,
\end{equation}
\begin{equation}
\left\langle \phi_{q12}\right\rangle \sim\mathcal{O}(50\,\mathrm{TeV})\,,
\end{equation}
\begin{equation}
M_{Q_{i}}\sim\mathcal{O}(100\,\mathrm{TeV})\,,
\end{equation}
\begin{equation}
M_{H_{2}^{u,d}}\sim\mathcal{O}(100\,\mathrm{TeV})\,,
\end{equation}
\begin{equation}
M_{H_{1}^{u,d}}\sim\mathcal{O}(10^{4}\,\mathrm{TeV})\,.
\end{equation}

\subsection{Neutrino masses and mixing\label{subsec:Neutrinos}}
Explaining the observed pattern of neutrino mixing and mass splittings
in gauge non-universal theories of flavour is usually difficult, due to the
accidental $U(2)^{5}$ flavour symmetry predicted by these models,
which is naively present in the neutrino sector as well. However,
exotic variations of the type I seesaw mechanism have been shown to
be successful in accommodating neutrino observations within non-universal
theories of flavour, see Refs.~\cite{FernandezNavarro:2023rhv,Fuentes-Martin:2020pww}.
Here we will incorporate the mechanism of \cite{FernandezNavarro:2023rhv},
which consists of adding SM singlet neutrinos which carry family hypercharges
(although their sum must of course vanish). These neutrinos can be seen as the fermionic counterpart of hyperons,
as they will connect the different
hypercharge sites, therefore breaking the $U(2)^{5}$ flavour symmetry in the neutrino sector. In this manner, these neutrinos allow to write effective operators which
may provide a successful pattern for neutrino mixing. However, the particular
model presented in \cite{FernandezNavarro:2023rhv} incorporates SM
singlet neutrinos with 1/4 family hypercharge factors, which cannot
be obtained from $SU(5)^{3}$, at least not from representations
with dimension smaller than $\mathbf{45}$~\footnote{Since these singlet neutrinos
can be seen as the fermionic counterpart of hyperons, the search for $SU(5)^{3}$ hyperon embeddings shown in Appendix~\ref{sec:Hyperons} shows that no neutrinos with 1/4 family hypercharge factors are found from $SU(5)^{3}$ representations with dimension up to $\mathbf{45}$.} according to a search with \texttt{GroupMath} \cite{Fonseca:2020vke}.

Following the recipe of Ref.~\cite{FernandezNavarro:2023rhv}, we
start by introducing two right-handed neutrinos: $N_{\mathrm{atm}}\sim(\mathbf{1,1})_{(0,2/3,-2/3)}$
and $N_{\mathrm{sol}}\sim(\mathbf{1,1})_{(2/3,0,-2/3)}$, which will
be responsible for atmospheric and solar neutrino mixing, respectively.
These neutrinos are embedded in $\Sigma_{23}\sim(\mathbf{1,10,\overline{10}})$ and $\Sigma_{13}\sim(\mathbf{10,1,\overline{10}})$
representations of $SU(5)^{3}$, respectively.
We also need to add the cyclic permutation $N_{\mathrm{cyclic}}$ embedded in $\Sigma_{12}\sim(\mathbf{10,\overline{10},1})$
to preserve the cyclic symmetry of $SU(5)^{3}$. However, we find that if the ``cyclic'' neutrino contained in $\Sigma_{12}$  is much heavier than the other neutrinos,
then we can ignore it as it decouples from the seesaw, and we recover the minimal framework of Ref.~\cite{FernandezNavarro:2023rhv}. Finally, in order to cancel gauge anomalies, we choose to make these neutrinos vector-like
by introducing the three corresponding conjugate neutrinos.

The next step is adding hyperons that provide effective Yukawa couplings
and Majorana masses for the singlet neutrinos. These are summarised
in the Dirac and Majorana mass matrices that follow (ignoring the
$\mathcal{O}(1)$ dimensionless couplings and the much heavier cyclic neutrinos)
\begin{flalign}
 &  &  & m_{D_{L}}=\frac{1}{M_\xi}\left(
\global\long\def\arraystretch{0.7}%
\begin{array}{@{}llc@{}}
 & \multicolumn{1}{c@{}}{\phantom{\!\,}\overline{N}_{\mathrm{sol}}} & \phantom{\!\,}\overline{N}_{\mathrm{atm}}\\
\cmidrule(l){2-3}\left.L_{1}\right| & 0 & 0\\
\left.L_{2}\right| & 0 & 0\\
\left.L_{3}\right| & \widetilde{\phi}_{u31}^{(-\frac{2}{3},0,\frac{2}{3})} & \widetilde{\phi}_{u32}^{(0,-\frac{2}{3},\frac{2}{3})}
\end{array}\right)H_{u}\,, &  & m_{D_{R}}=\frac{1}{M_\xi}\left(
\global\long\def\arraystretch{0.7}%
\begin{array}{@{}llc@{}}
 & \multicolumn{1}{c@{}}{\phantom{\!\,}N_{\mathrm{sol}}} & \phantom{\!\,}N_{\mathrm{atm}}\\
\cmidrule(l){2-3}\left.L_{1}\right| & \phi_{q13}^{(-\frac{1}{6},0,\frac{1}{6})} & 0\\
\left.L_{2}\right| & \phi^{(-\frac{2}{3},\frac{1}{2},\frac{1}{6})}_{u123} & \phi_{q23}^{(0,-\frac{1}{6},\frac{1}{6})}\\
\left.L_{3}\right| & \phi_{u31}^{(-\frac{2}{3},0,\frac{2}{3})} & \phi_{u32}^{(0,-\frac{2}{3},\frac{2}{3})}
\end{array}\right)H_{u}\,,\label{eq:mDL_mDR}\\
\nonumber \\
 &  &  & M_{L}=\left(
\global\long\def\arraystretch{0.7}%
\begin{array}{@{}llc@{}}
 & \multicolumn{1}{c@{}}{\phantom{\!\,}\overline{N}_{\mathrm{sol}}} & \phantom{\!\,}\overline{N}_{\mathrm{atm}}\\
\cmidrule(l){2-3}\left.\:\:\overline{N}_{\mathrm{sol}}\right| & \tilde{\phi}_{\mathrm{sol}}^{(-\frac{4}{3},0,\frac{4}{3})} & 0\\
\left.\overline{N}_{\mathrm{atm}}\right| & 0 & \widetilde{\phi}_{\mathrm{atm}}^{(0,-\frac{4}{3},\frac{4}{3})}
\end{array}\right)\,, &  & M_{R}=\left(
\global\long\def\arraystretch{0.7}%
\begin{array}{@{}llc@{}}
 & \multicolumn{1}{c@{}}{\phantom{\!\,}N_{\mathrm{sol}}} & \phantom{\!\,}N_{\mathrm{atm}}\\
\cmidrule(l){2-3}\left.\:\:N_{\mathrm{sol}}\right| & \phi_{\mathrm{sol}}^{(-\frac{4}{3},0,\frac{4}{3})} & 0\\
\left.N_{\mathrm{atm}}\right| & 0 & \phi_{\mathrm{atm}}^{(0,-\frac{4}{3},\frac{4}{3})}
\end{array}\right)\,,\label{eq:ML_MR}
\end{flalign}
\begin{equation}
M_{LR}=\left(
\global\long\def\arraystretch{0.7}%
\begin{array}{@{}llc@{}}
 & \multicolumn{1}{c@{}}{\phantom{\!\,}N_{\mathrm{sol}}} & \phantom{\!\,}N_{\mathrm{atm}}\\
\cmidrule(l){2-3}\left.\:\:\overline{N}_{\mathrm{sol}}\right| & M_{N_{\mathrm{sol}}} & 0\\
\left.\overline{N}_{\mathrm{atm}}\right| & 0 & M_{N_{\mathrm{atm}}}
\end{array}\right)\,,\label{eq:MLR}
\end{equation}
\begin{figure}[t]
\subfloat[]{\resizebox{.32\textwidth}{!}{
\begin{tikzpicture}
	\begin{feynman}
		\vertex (a);
		\vertex [right=16mm of a] (b);
		\vertex [right=of b] (c);
		\vertex [right=of c] (d);
		\vertex [right=of d] (e);
		\vertex [above=of b] (f1) {\(H^{u}_{3}\)};
		\vertex [above=of d] (f2) {\(\phi_{u32}{,}\,\phi_{u31}\)};
		\diagram* {
			(a) -- [fermion, edge label'=\(\ell_{3}\)] (b) -- [charged scalar] (f1),
			(b) -- [edge label'=\(\xi_0\)] (c),
			(c) -- [edge label'=\(\xi_0\), insertion=0] (d) -- [charged scalar] (f2),
			(d) -- [fermion, edge label'=\(N_{\mathrm{atm}}{,}\,N_{\mathrm{sol}}\)] (e),
	};
	\end{feynman}
\end{tikzpicture}
}

}\subfloat[]{\begin{centering}
\resizebox{.32\textwidth}{!}{ \begin{tikzpicture}
	\begin{feynman}
		\vertex (a);
		\vertex [right=16mm of a] (b);
		\vertex [right=of b] (c);
		\vertex [right=of c] (d);
		\vertex [right=of d] (e);
		\vertex [above=of b] (f1) {\(H^{u}_{3}\)};
		\vertex [above=of d] (f2) {\(\phi_{q23}{,}\,\phi_{u123}\)};
		\diagram* {
			(a) -- [fermion, edge label'=\(\ell_{2}\)] (b) -- [charged scalar] (f1),
			(b) -- [edge label'=\(\xi_{23}\)] (c),
			(c) -- [edge label'=\(\xi_{23}\), insertion=0] (d) -- [charged scalar] (f2),
			(d) -- [fermion, edge label'=\(N_{\mathrm{atm}}{,}\,N_{\mathrm{sol}}\)] (e),
	};
	\end{feynman}
\end{tikzpicture}}
\par\end{centering}
}\subfloat[]{\begin{centering}
\resizebox{.32\textwidth}{!}{ \begin{tikzpicture}
	\begin{feynman}
		\vertex (a);
		\vertex [right=16mm of a] (b);
		\vertex [right=of b] (c);
		\vertex [right=of c] (d);
		\vertex [right=of d] (e);
		\vertex [above=of b] (f1) {\(H^{u}_{3}\)};
		\vertex [above=of d] (f2) {\(\phi_{q13}\)};
		\diagram* {
			(a) -- [fermion, edge label'=\(\ell_{1}\)] (b) -- [charged scalar] (f1),
			(b) -- [edge label'=\(\xi_{13}\)] (c),
			(c) -- [edge label'=\(\xi_{13}\), insertion=0] (d) -- [charged scalar] (f2),
			(d) -- [fermion, edge label'=\(N_{\mathrm{sol}}\)] (e),
	};
	\end{feynman}
\end{tikzpicture}}
\par\end{centering}
}\caption[]{Diagrams leading to effective Yukawa couplings in the neutrino sector. \label{fig:Diagrams_neutrinos}}
\end{figure}
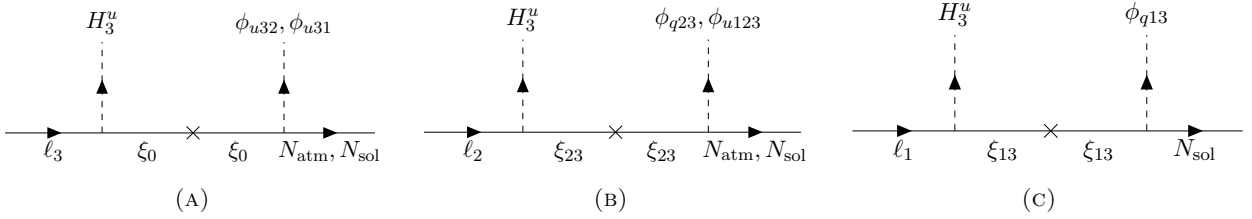
where the heavy scale $M_\xi$ is associated to the mass of the heavy vector-like fermions $\xi_0\sim(\mathbf{1,1})_{(0,0,0)}$, $\xi_{23}\sim(\mathbf{1,1})_{(0,1/2,-1/2)}$
(plus cyclic permutations), which are embedded in the representations
$\Xi_{0}\sim(\mathbf{1,1,1})$ and $\Xi_{23}\sim(\mathbf{1,5,\overline{5}})$ (plus conjugate, plus cyclic
permutations) of $SU(5)^{3}$. Example diagrams are shown in Fig~\ref{fig:Diagrams_neutrinos}.
We now construct the full neutrino mass matrix as
\begin{equation}
M_{\nu}=\left(
\global\long\def\arraystretch{0.7}%
\begin{array}{@{}llcc@{}}
 & \multicolumn{1}{c@{}}{\phantom{\!\,}\nu} & \phantom{\!\,}\overline{N} & \phantom{\!\,}N\\
\cmidrule(l){2-4}\left.\:\,\nu\right| & 0 & m_{D_{L}} & m_{D_{R}}\\
\left.\overline{N}\right| & m_{D_{L}}^{\mathrm{T}} & M_{L} & M_{LR}\\
\left.N\right| & m_{D_{R}}^{\mathrm{T}} & M_{LR}^{\mathrm{T}} & M_{R}
\end{array}\right)\equiv\left(\begin{array}{cc}
0 & m_{D}\\
m_{D}^{\mathrm{T}} & M_{N}
\end{array}\right)\,,\label{eq:Full_Mnu}
\end{equation}
where we have defined $\nu$ as a 3-component vector containing the
weak eigenstates of active neutrinos, while $N$ and $\overline{N}$
are 2-component vectors containing the SM singlets $N$ and conjugate
neutrinos $\overline{N}$ , respectively. Now we assume that all the
hyperons in Eqs.~(\ref{eq:ML_MR}-\ref{eq:MLR}) get VEVs at the
scale $v_{23}$ of 23 hypercharge breaking according to Eq.~\eqref{eq:23_breaking},
and we have into account that $\left\langle \phi_{q13}\right\rangle /\left\langle \phi_{q23}\right\rangle \approx\lambda$
as obtained from the discussion of the charged fermion sector in Section~\ref{subsec:Charged_fermions}.
It is also required to assume $M_{N_{\mathrm{atm}}},\,M_{N_{\mathrm{atm}}}\apprle v_{23}$
in order to obtain the observed neutrino mixing with the textures of Eqs.~(\ref{eq:ML_MR}-\ref{eq:MLR}).

Dirac-type masses in $m_{D_{L,R}}$ may be orders of magnitude smaller
than the electroweak scale, because they arise from non-renormalisable
operators proportional to the SM VEV. In contrast, the eigenvalues
of $M_{N}$ are not smaller than $\mathcal{O}(v_{23})$, which is
at least TeV. Therefore, the condition $m_{D}\ll M_{N}$ is fulfilled
in Eq.~\eqref{eq:Full_Mnu} and we can safely apply the seesaw formula
to obtain, up to $\mathcal{O}(1)$ factors,
\begin{equation}
m_{\nu}\simeq m_{D}M_{N}^{-1}m_{D}^{\mathrm{T}}\approx\left(\begin{array}{ccc}
1 & 1 & \lambda\\
1 & 1 & 1\\
\lambda & 1 & 1
\end{array}\right)v_{23}\frac{v_{\mathrm{SM}}^{2}}{M_\xi^{2}}\,.\label{eq:Neutrino_texture}
\end{equation}
This is the same texture that was obtained in Ref.~\cite{FernandezNavarro:2023rhv},
which is able to accommodate all the observed neutrino mixing angles
and mass splittings \cite{deSalas:2020pgw,Gonzalez-Garcia:2021dve}
with $\mathcal{O}(1)$ parameters once the dimensionless coefficients
implicit in Eq.~\eqref{eq:Neutrino_texture} are considered. Remarkably,
the singlet neutrinos $N_{\mathrm{atm}}$ and $N_{\mathrm{sol}}$
get masses around the TeV scale $(v_{23})$ and contribute to the
RGE, while the cyclic neutrino is assumed to get a very heavy vector-like
mass and decouples, as mentioned before.

\subsection{Gauge coupling unification} \label{sec:gcu}

In order to ensure that the gauge couplings of our model do indeed unify into a single value at some high energy scale, we must solve their one-loop RGEs, which take the generic form~\cite{Machacek:1983tz}
\begin{equation}
\frac{d g_i}{d \ln \mu} = \frac{g_i^3}{16 \, \pi^2} \, b_i \, .
\end{equation}
The $b_i$ coefficients depend on the specific group $G_i$, with gauge coupling $g_i$, and the representations in the model. They are given by
\begin{equation}
    b_i = -\frac{11}{3} \, C_2(G_i) + \frac{4}{3} \, \kappa \, S_2(F_i) + \frac{1}{6} \, \eta \, S_2(S_i) \, .
\end{equation}
Here $\mu$ is the renormalization scale, $C_2(G_i)$ is the quadratic Casimir of the adjoint representation of $G_i$ and $S_2(F_i)$ and $S_2(S_i)$ are the sums of the Dynkin indices of all fermion and scalar non-trivial representations under $G_i$. Finally, $\kappa = 1 \, (1/2)$ for Dirac (Weyl) fermions and $\eta = 2 \, (1)$ for complex (real) scalars.

\begin{table}[tb]
\centering{}%
\begin{tabular}{ccc}
\toprule 
\textbf{Regime} & \textbf{Gauge group} & \textbf{$\boldsymbol{b_i}$ coefficients} \tabularnewline
\midrule 
1 & SM$^3$ & $\left( -\frac{22}{3}, -3, \frac{46}{15}, -\frac{22}{3}, -3, \frac{46}{15}, -\frac{22}{3}, -3, \frac{46}{15} \right)$ \\
2 & $SU(3)_c \times SU(2)_L \times U(1)_{Y_1} \times U(1)_{Y_2} \times U(1)_{Y_3}$ & $\left( 0, \frac{11}{3}, \frac{122}{45}, \frac{122}{45}, \frac{46}{15} \right)$ \\
3 & $SU(3)_c \times SU(2)_L \times U(1)_{Y_1} \times U(1)_{Y_2} \times U(1)_{Y_3}$ & $\left( 0, \frac{11}{3}, \frac{104}{45}, \frac{104}{45}, \frac{8}{3} \right)$ \\
4 & $SU(3)_c \times SU(2)_L \times U(1)_{Y_1} \times U(1)_{Y_2} \times U(1)_{Y_3}$ & $\left( 0, \frac{10}{3}, \frac{19}{9}, \frac{104}{45}, \frac{8}{3} \right)$ \\
5 & $SU(3)_c \times SU(2)_L \times U(1)_{Y_1} \times U(1)_{Y_2} \times U(1)_{Y_3}$ & $\left( 0, 3, \frac{19}{9}, \frac{19}{9}, \frac{8}{3} \right)$ \\
6 & $SU(3)_c \times SU(2)_L \times U(1)_{Y_1} \times U(1)_{Y_2} \times U(1)_{Y_3}$ & $\left( -4, -3, \frac{89}{45}, \frac{89}{45}, \frac{38}{15} \right)$ \\
7 & $SU(3)_c \times SU(2)_L \times U(1)_{Y_{12}} \times U(1)_{Y_3}$ & $\left( -7, -3, \frac{11}{3},\frac{38}{15} \right)$ \\
8 & $SU(3)_c \times SU(2)_L \times U(1)_Y$ & $\left( -7, -3, \frac{21}{5} \right)$ \\
\bottomrule
\end{tabular}\caption{$b_i$ coefficients of our model. See Appendix~\ref{sec:app} for details on the gauge symmetries and particle content at each energy regime. \label{tab:bcoeffs}}
\end{table}

\begin{figure}[tb!]
  \centering
  \includegraphics[width=0.75\linewidth]{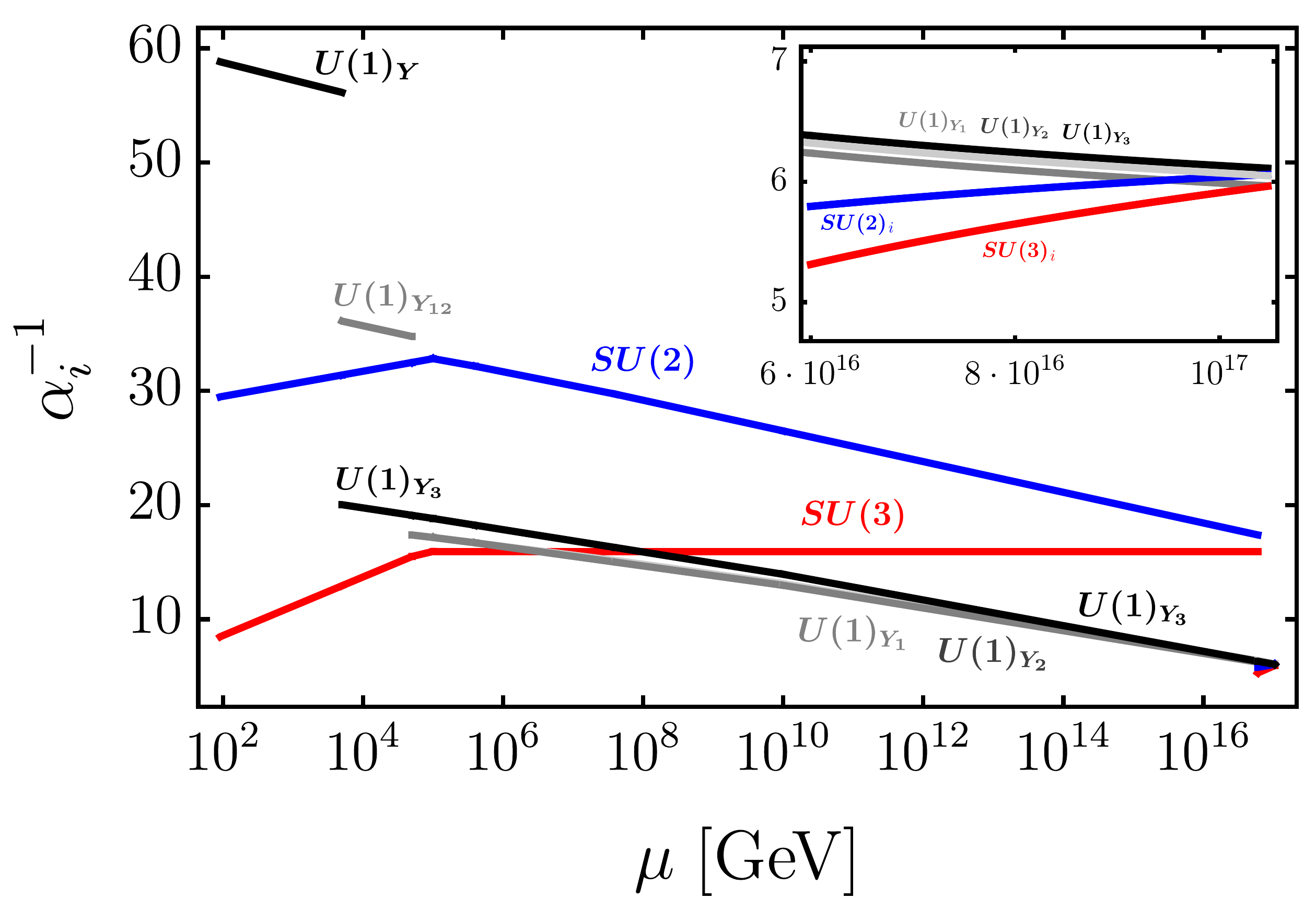}
  \caption{Running of the gauge couplings. The red lines correspond to the $SU(3)$ gauge couplings, the blue ones to the $SU(2)$ gauge couplings and the black/grey ones to the $U(1)$ gauge couplings. A zoom-in with the high-energy region close to the unification scale is also shown. These results have been obtained with $v_{23} = 5$ TeV, $v_{12} = 50$ TeV, $M_Q = 100$ TeV, $M_{H_2^{u,d}} = 400$ TeV, $M_{H_1^{u,d}} = 4 \cdot 10^{4}$ TeV, $M_\xi = 10^{10}$ GeV and $v_{\rm{SM}^3} = 6 \cdot 10^{16}$ GeV. The discontinuities in the plot are due to the gauge coupling matching conditions that apply at each symmetry breaking step, see main text and Appendix~\ref{sec:app}.
    \label{fig:running}
    }
\end{figure}

We computed the $b_i$ coefficients of our model, taking into account not only the gauge group for each energy regime, but also the particle content, since a particle decouples and does not contribute to the running at energies below its mass. The gauge symmetries and particle content at each energy regime are described in detail in Appendix~\ref{sec:app}, whereas our results for the $b_i$ coefficients of the model are given in Table~\ref{tab:bcoeffs}. Finally, we display results for the running of the gauge couplings in Fig.~\ref{fig:running}. This figure has been obtained by fixing the intermediate energy scales to
\begin{equation} \label{eq:scales}
  \begin{split}
    &v_{23} = 5 \; \text{TeV} \, , \\   
    &M_{\Theta} = 100 \; \text{TeV} \, , \\
    &M_\xi = 10^{10} \; \text{GeV} \, , 
  \end{split}
\quad
  \begin{split}
    &v_{12} = 50 \, \text{TeV} \; , \\
    &M_{H_2^{u,d}} = 400 \, \mathrm{TeV} , \\
    &v_{\rm{SM}^3} = 6 \cdot 10^{16} \; \text{GeV} \,.
  \end{split}
  \quad
  \begin{split}
   &M_Q = 100 \; \text{TeV} \; , \\
   &M_{H_1^{u,d}} = 4 \cdot 10^4 \; \text{TeV} \; , \\
   &
  \end{split}
\end{equation}
Before discussing gauge coupling unification, we note that we expect radiative corrections to disturb the scales above unless some couplings in the scalar potential are fine-tuned to some extent. In particular, the bi-adjoint scalars $\Omega_{ij}$ may couple at tree-level to the light Higgs doublets in order to give corrections to their masses proportional to the GUT scale. This is a consequence of the well-known \textit{hierarchy problem} that afflicts all non-supersymmetric GUTs. To avoid this, perhaps one could build a $SU(5)^{3}$ tri-unification supersymmetric theory, but this would lead to rapid proton decay via $d=5$ operators mediated by the coloured Higgs triplet superpartners\footnote{In standard supersymmetric $SU(5)$ this contribution also exists but is suppressed by small Yukawa couplings connected to first family fermion masses. In contrast, in $SU(5)^{3}$ tri-unification all Yukawa couplings are $\mathcal{O}(1)$, including those of the coloured Higgs, hence rapid proton decay happens even if we push the masses of coloured Higgs superpartners to the Planck scale.}. A solution to this would require further model building \cite{Barbieri:1994cx,Babu:2007mb}. Another option would be to include a strongly coupled sector to generate the light scalars as pseudo Goldstone bosons, e.g.~one could introduce a strongly coupled $SU(5)$ \cite{Georgi:1981gj} and generalise the permutation symmetry to $\mathbb{Z}_{4}$ in order to enforce a single gauge coupling. This would require further model building beyond the scope of this paper.

\begin{figure}[tb!]
  \centering
  \includegraphics[width=0.49\linewidth]{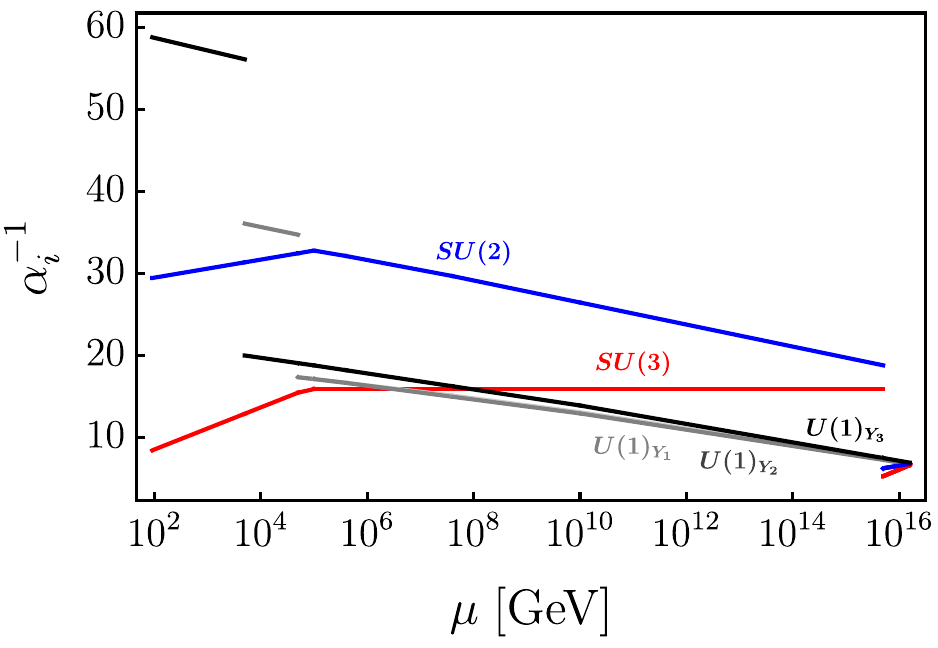}
  \includegraphics[width=0.49\linewidth]{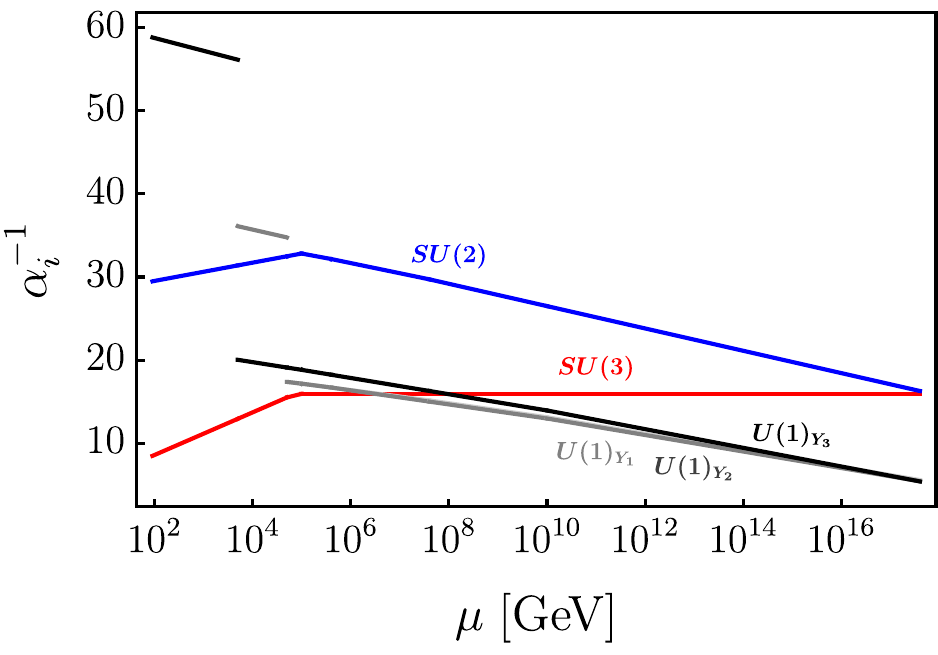}
  \caption{Running of the gauge couplings. Colour code as in Fig.~\ref{fig:running}. These results have been obtained with $v_{\rm{SM}^3} = 5 \cdot 10^{15}$ GeV (left) and $v_{\rm{SM}^3} = 5 \cdot 10^{17}$ GeV (right). The rest of the intermediate scales have been chosen as in Eq.~\eqref{eq:scales}. The discontinuities in the plots are due to the gauge coupling matching conditions that apply at each symmetry breaking step, see main text and Appendix~\ref{sec:app}
    \label{fig:running2}
    }
\end{figure}

Nevertheless, without the need of imposing supersymmetry nor extra dynamics, the nine gauge couplings of the SM$^3$ group unify at a very high unification scale $M_{\rm GUT} \approx 10^{17}$ GeV, slightly above the SM$^3$ breaking scale, with a unified gauge coupling $g_{\rm GUT} \approx 1.44$. We note the important role played by three $\Theta_{i}$ colour
octets embedded into $\Omega_{ij}$, and by the $Q_{i}$ vector-like quarks which also act as heavy messengers of the flavour theory,
which are crucial to modify the running of the $SU(3)$ and $SU(2)$ gauge couplings
in order to achieve unification. We also remark that the discontinuities in Figs.~\ref{fig:running} and \ref{fig:running2} are due to the gauge coupling matching conditions that apply at the steps in which the $U(1)_{Y}$ group is decomposed into two (first discontinuity) and three hypercharges (second discontinuity) and in which the $SU(3)$ and $SU(2)$ groups are decomposed into one for each family (third discontinuity), see more details in Appendix~\ref{sec:app}.

Even though the $\mathbb{Z}_{3}$ symmetry gets broken at the SM$^3$ breaking scale, it stays approximate at low energies, down to the tri-hypercharge breaking scale, and only the running of $U(1)_{Y_3}$ is slightly different from that of the other two hypercharge groups. In fact, the gauge couplings of the $U(1)_{Y_1}$ and $U(1)_{Y_2}$ groups almost overlap and cannot be distinguished in Fig.~\ref{fig:running}. This can be easily understood by inspecting the $b_i$ coefficients on Table~\ref{tab:bcoeffs}. Then, the matching conditions at $v_{12} = 50$ TeV split the low energy $g_{Y_{12}}$ and $g_{Y_3}$ couplings, which become clearly different: $g_{Y_{12}}(v_{12}) \approx 0.59$ and $g_{Y_3}(v_{12}) \approx 0.79$. Finally, at $v_{23} = 5$ TeV one recovers the standard $SU(3)_c \times SU(2)_L \times U(1)_Y$ gauge group, which remains unbroken down to the electroweak scale.

\begin{figure}
\centering
\begin{minipage}{0.53\textwidth}
  \centering
  \includegraphics[width=\linewidth]{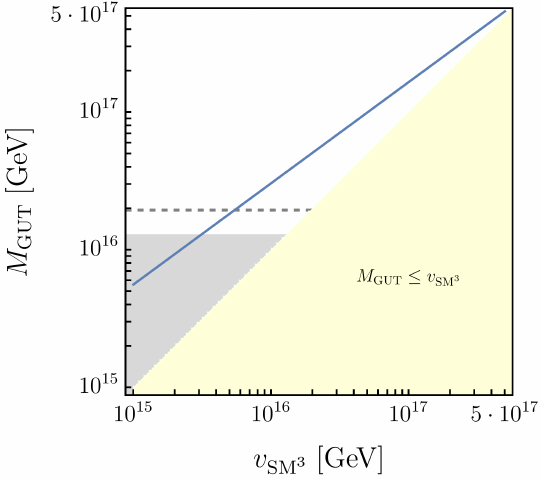}
\end{minipage}%
\begin{minipage}{0.445\textwidth}
  \centering
  \vspace*{-0.2cm}
  \includegraphics[width=\linewidth]{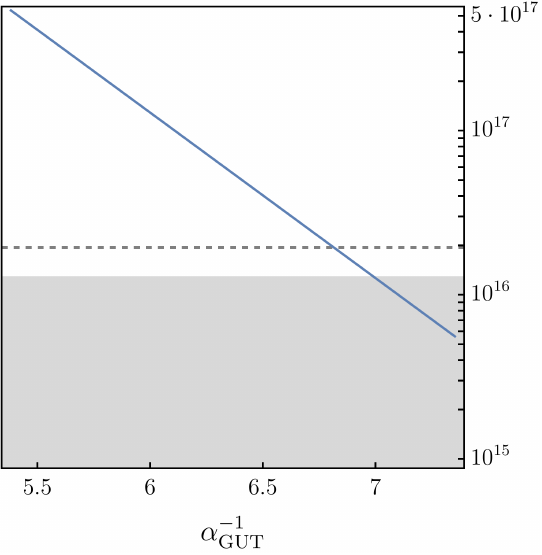}
\end{minipage}
\caption{$M_{\rm GUT}$ as a function of $v_{\rm{SM}^3}$ (left) and $\alpha_{\rm GUT}^{-1}$ (right). The SM$^3$ breaking scale $v_{\rm{SM}^3}$ varies in these plots, while the rest of the intermediate scales have been fixed to the values in Eq.~\eqref{eq:scales}. The shaded grey region is excluded by the existing Super-Kamiokande 90\% C.L. limit on the $p \to e^+ \pi^0$ lifetime, $\tau(p \to e^+ \pi^0) > 2.4 \cdot 10^{34}$ years~\cite{Super-Kamiokande:2020wjk}, whereas the horizontal dashed line corresponds to the projected Hyper-Kamiokande sensitivity at 90\% C.L. after 20 years of runtime, $\tau(p \to e^+ \pi^0) > 1.2 \cdot 10^{35}$ years, obtained in~\cite{Bhattiprolu:2022xhm}. See Section~\ref{subsec:protondecay} for details on the proton decay calculation. Finally, the shaded yellow region on the left-hand plot is excluded due to $M_{\rm GUT} \leq v_{\rm{SM}^3}$.
    \label{fig:running3}
    }
\end{figure}

In order to study how unification changes with the scale of SM$^3$ breaking, $v_{\rm{SM}^3}$, we consider the values $v_{\rm{SM}^3} = 5 \cdot 10^{15}$ GeV and $v_{\rm{SM}^3} = 5 \cdot 10^{17}$ GeV and fix the rest of the intermediate scales as in Eq.~\eqref{eq:scales}. Results for the running of the gauge couplings in these two scenarios are shown in Fig.~\ref{fig:running2}. On the left-hand side we show the case $v_{\rm{SM}^3} = 5 \cdot 10^{15}$ GeV whereas on the right-hand side we display our results for $v_{\rm{SM}^3} = 5 \cdot 10^{17}$ GeV. In the first case, our choice of SM$^3$ breaking scale leads to unification of the gauge couplings at a relatively low scale, $M_{\rm GUT} \approx 1.8 \cdot 10^{16}$ GeV. This is potentially troublesome, as it may lead to too fast proton decay, as explained below. In contrast, when the SM$^3$ breaking scale is chosen to be very high, as in the second scenario, unification also gets delayed to much higher energies. In fact, we note that with our choice $v_{\rm{SM}^3} = 5 \cdot 10^{17}$ GeV, gauge coupling unification already takes place at the SM$^3$ breaking scale, $M_{\rm GUT} \approx v_{\rm{SM}^3}$. In this case, $SU(5)^3$ breaks \textit{directly} to the tri-hypercharge group $SU(3)_c \times SU(2)_L \times U(1)_{Y_1} \times U(1)_{Y_2} \times U(1)_{Y_3}$ and there is no intermediate SM$^3$ scale. Finally, the impact of $v_{\rm{SM}^3}$ is further illustrated in Fig.~\ref{fig:running3}. Here we show the relation between $M_{\rm GUT}$, $v_{\rm{SM}^3}$ and $\alpha_{\rm GUT}^{-1}=4\pi/g^{2}_{\rm GUT}$. These two plots have been made by varying $v_{\rm{SM}^3}$ and all the other intermediate scales fixed as in Eq.~\eqref{eq:scales}. The left-hand side of this figure confirms that larger $v_{\rm{SM}^3}$ values lead to higher unification scales and smaller gaps between these two energy scales. The right-hand side of the figure shows the relation between the unified gauge coupling and the GUT scale. Again, the larger $M_{\rm GUT}$ (or, equivalently, larger $v_{\rm{SM}^3}$) is, the larger $g_{\rm GUT}$ (and smaller $\alpha_{\rm GUT}^{-1}$) becomes. In particular, in this plot $g_{\rm GUT}$ ranges from $\sim 1.30$ to $\sim 1.53$.

\subsection{Proton decay}
\label{subsec:protondecay}

As in any GUT, proton decay is a major prediction in our setup. In standard $\GGM$ the most relevant proton decay mode is usually $p \to e^+ \pi^0$. This process is induced by the tree-level exchange of the $X$ gauge bosons contained in the \24 (adjoint) representation, such as the $(\three,\two)_{\frac{5}{6}}$ vector leptoquark. Integrating out these heavy vector leptoquarks leads to effective dimension-6 operators\footnote{We note that a coloured Higgs triplet in our model also contributes to dimension-6 proton decay operator, with a similar size as the gauge leptoquark contribution, hence just changing the Wilson coefficients by $\mathcal{O}(1)$ factors.} that violate both baryon and lepton
numbers, for instance $qqq\ell$. The resulting proton life time can be roughly estimated as
\begin{equation} \label{eq:proton_app}
  \tau_p \approx \tau_p^{\rm app} = \frac{m_X^4}{\alpha_{\rm GUT}^2 m_p^5} \, ,
\end{equation}
where $m_X$ is the mass of the heavy leptoquark, $m_p \approx 0.938$ GeV is the proton mass and $\alpha_{\rm GUT} = g_{\rm GUT}^2/(4 \pi)$ is the value of the fine structure constant at the unification scale. For a comprehensive review on proton decay we refer to~\cite{Nath:2006ut}.

In our model there are three $\GGM$ groups. This implies a larger
number (three times as many) of vector leptoquarks, potentially
affecting the proton lifetime. However, due the special flavour
structure of our setup, only one of the leptoquark \textit{generations}
couples directly to the first fermion generation. The other two couple
to the first SM fermion generation only via mixing. However, given that in our setup the three generation
leptoquarks get the same mass, in practice the gauge leptoquark phenomenology is that of
conventional (flavour universal) $SU(5)$.
One can easily estimate that for $m_X = 10^{17}$ GeV and $g_{\rm GUT} \sim 1.5$, the
proton life time is $\tau_p \sim 10^{38}$ years, well above the
current experimental limit, $\tau(p \to e^+ \pi^0) > 2.4 \cdot
10^{34}$ years at 90\% C.L. ~\cite{Super-Kamiokande:2020wjk}. Therefore, a large
unification scale suffices to guarantee that our model respects the
current limits on the proton lifetime. In fact, such a long life time is beyond the reach of near future experiments, which will increase the current limit by about one order of magnitude~\cite{Bhattiprolu:2022xhm}.

A more precise determination of the $p \to e^+ \pi^0$ decay width is \cite{Nath:2006ut,Chakrabortty:2019fov}
\begin{flalign}
\Gamma(p\rightarrow e^{+}\pi^{0})= & \frac{m_{p}}{8}\pi\left(1-\frac{m_{\pi^{0}}^{2}}{m_{p}^{2}}\right)^{2}A_{L}^{2}\frac{\alpha_{\mathrm{GUT}}^{2}}{M_{\mathrm{GUT}}^{4}}\times\left[A_{SL}^{2}\left|C_{L}\right|^{2}\left|\langle \pi^{0}|(ud)_{L}u_{L}|p\rangle \right|^{2}\right.\label{eq:proton_decay_width}\\
 & \left.+A_{SR}^{2}\left|C_{R}\right|^{2}\left|\langle \pi^{0}|(ud)_{R}u_{L}|p\rangle \right|^{2}\right]\nonumber 
\end{flalign}
where $A_{L}\approx1.247$ accounts for the QCD RGE from the $M_{Z}$
scale to $m_{p}$ \cite{Nath:2006ut}. In contrast, $A_{SL(R)}$ accounts for the short-distance
RGE from the GUT scale to $M_{Z}$, given by 
\begin{table}[tb]
\centering{}%
\begin{tabular}{ccc}
\toprule 
\textbf{Gauge group}  & \textbf{$\gamma_{iL}$ coefficients}  & \textbf{$\gamma_{iR}$ coefficients} \tabularnewline
\midrule 
SM$^{3}$  & $\left(2,\frac{9}{4},\frac{23}{20},0,0,0,0,0,0\right)$  & $\left(2,\frac{9}{4},\frac{11}{20},0,0,0,0,0,0\right)$\tabularnewline
$SU(3)_{c}\times SU(2)_{L}\times U(1)_{Y_{1}}\times U(1)_{Y_{2}}\times U(1)_{Y_{3}}$  & $\left(2,\frac{9}{4},\frac{23}{20},0,0\right)$ & $\left(2,\frac{9}{4},\frac{11}{20},0,0\right)$\tabularnewline
$SU(3)_{c}\times SU(2)_{L}\times U(1)_{Y_{12}}\times U(1)_{Y_{3}}$  & $\left(2,\frac{9}{4},\frac{23}{20},0\right)$ & $\left(2,\frac{9}{4},\frac{11}{20},0\right)$\tabularnewline
$SU(3)_{c}\times SU(2)_{L}\times U(1)_{Y}$  & $\left(2,\frac{9}{4},\frac{23}{20}\right)$ & $\left(2,\frac{9}{4},\frac{11}{20}\right)$\tabularnewline
\bottomrule
\end{tabular}\caption{Anomalous dimension coefficients $\gamma_{iL,R}$ for proton decay operators in our model. \label{tab:gammacoeffs}}
\end{table}
\begin{equation}
A_{SL(R)}=\prod_{A}^{M_{Z}\leq M_{A}\leq M_{\mathrm{GUT}}}\prod_{i}\left[\frac{\alpha_{i}(M_{A+1})}{\alpha_{i}(M_{A})}\right]^{\frac{\gamma_{iL(R)}}{b_{i}}}\,,
\end{equation}
where $b_{i}$ and $\gamma_{i}$ denote the $\beta$ coefficients
and the anomalous dimensions computed at one-loop in Tables \ref{tab:bcoeffs} and \ref{tab:gammacoeffs}, respectively,
for the various intermediate scales $M_{A}$. The matrix elements
appearing in Eq.~(\ref{eq:proton_decay_width}) are given by \cite{Aoki:2017puj}
\begin{equation}
\langle \pi^{0}|(ud)_{L}u_{L}|p\rangle =0.134(5)(16)\,\mathrm{GeV^{2}}\,,
\end{equation}
\begin{equation}
\langle \pi^{0}|(ud)_{R}u_{L}|p\rangle =-0.131(4)(13)\,\mathrm{GeV^{2}}\,,
\end{equation}
where the errors (shown in the parenthesis) denote statistical and systematic uncertainties, respectively. 
Given that in our model we have three $SU(5)$ groups, we actually have
three generations of the usual $SU(5)$ leptoquarks, coupling only to their
corresponding family of chiral fermions. However, since the three
$SU(5)_{i}$ groups are all broken down to their $\mathrm{SM}_{i}$ subgroups
at the same scale, in practice the model reproduces the phenomenology
of a flavour universal leptoquark as in conventional $SU(5),$ albeit
with the specific fermion mixing predicted by our model as shown in
Section~\ref{subsec:Charged_fermions}. The effect of fermion mixing is encoded via the coefficients
$C_{L}$ and $C_{R}$ \cite{Nath:2006ut}
\begin{equation}
C_{L}=(V_{u^{c}}^{\dagger}V_{u})^{11}(V_{e^{c}}^{\dagger}V_{d})^{11}+(V_{u^{c}}^{\dagger}V_{u}V_{\mathrm{CKM}})^{11}(V_{e^{c}}^{\dagger}V_{d}V_{\mathrm{CKM}}^{\dagger})^{11}\,,
\end{equation}
\begin{equation}
C_{R}=(V_{u^{c}}^{\dagger}V_{u})^{11}(V_{d^{c}}^{\dagger}V_{e})^{11}\,,
\end{equation}
where $V_{\mathrm{CKM}}=V_{u}^{\dagger}V_{d}$. Notice that even though our flavour model predicts non-generic fermion mixing, the alignment of the CKM matrix is not univocally predicted but relies on the choice of dimensionless coefficients. Assuming the CKM mixing to originate mostly from the down sector we find $C_L\simeq 1.946$ and $C_R\simeq 0.999$, while if the CKM mixing originates mostly from the up sector we find very similar coefficients as $C_L\simeq 1.946$ and $C_R\simeq 0.974$. Therefore, the prediction for proton decay is robust and independent of the alignment of the CKM to excellent accuracy.
\begin{figure}
\centering
\begin{minipage}{0.53\textwidth}
  \centering
  \includegraphics[width=\linewidth]{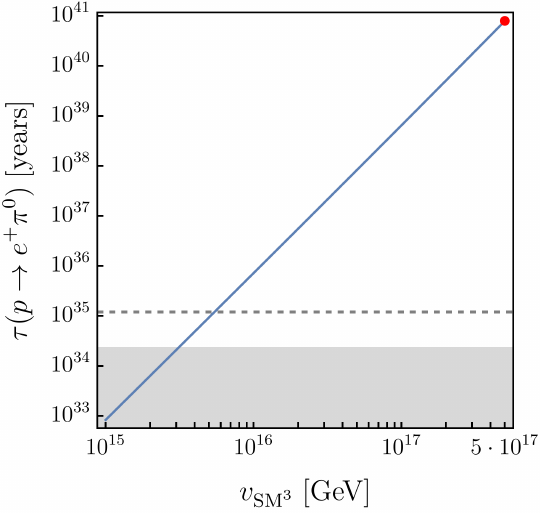}
\end{minipage}%
\begin{minipage}{0.445\textwidth}
  \centering
  \vspace*{-0.2cm}
  \includegraphics[width=\linewidth]{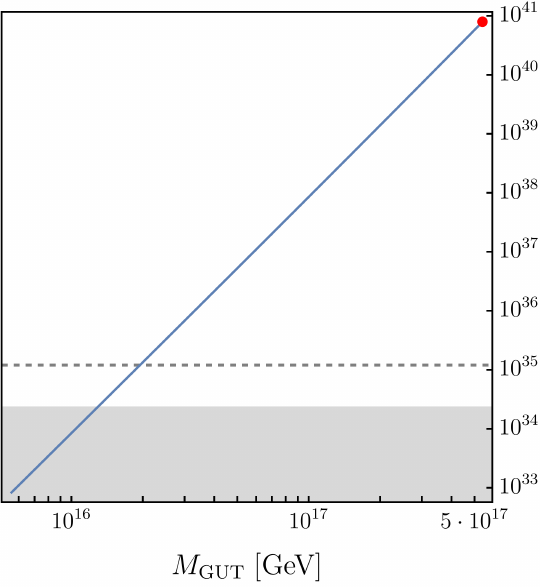}
\end{minipage}
\caption{$\tau(p \to e^+ \pi^0)$ as a function of $v_{\rm{SM}^3}$ (left) and $M_{\rm GUT}$ (right). The SM$^3$ breaking scale $v_{\rm{SM}^3}$ varies in these plots, while the rest of the intermediate scales have been fixed to the values in Eq.~\eqref{eq:scales}. The choice $v_{\rm{SM}^3} = 5 \cdot 10^{17}$ GeV, which leads to $M_{\rm GUT} \approx v_{\rm{SM}^3}$, is highlighted with a red point in both plots. The shaded grey region is excluded by the existing Super-Kamiokande 90\% C.L. limit on the $p \to e^+ \pi^0$ lifetime, $\tau(p \to e^+ \pi^0) > 2.4 \cdot 10^{34}$ years~\cite{Super-Kamiokande:2020wjk}, whereas the horizontal dashed line corresponds to the projected Hyper-Kamiokande sensitivity at 90\% C.L. after 20 years of runtime, $\tau(p \to e^+ \pi^0) > 1.2 \cdot 10^{35}$ years, obtained in~\cite{Bhattiprolu:2022xhm}.
    \label{fig:proton}
    }
\end{figure}

We show our numerical results for the $p \to e^+ \pi^0$ lifetime in Fig.~\ref{fig:proton}. Again, $v_{\rm{SM}^3}$ varies in the left panel of this figure, while the rest of intermediate scales have been chosen as in Eq.~\eqref{eq:scales}. The right panel shows an equivalent plot with the $p \to e^+ \pi^0$ lifetime as a function of $M_{\rm GUT}$. This figure provides complementary information to that already shown in Fig.~\ref{fig:running3}. In both cases we have used the precise determination of the lifetime in Eq.~\eqref{eq:proton_decay_width}, but we note that the estimate in Eq.~\eqref{eq:proton_app} actually provides a very good approximation, with $\tau_p / \tau_p^{\rm app} \in \left[0.5, 1.2 \right]$ in the parameter region covered in Fig.~\ref{fig:proton}. The current Super-Kamiokande 90\% C.L. limit on the $p \to e^+ \pi^0$ lifetime, $\tau(p \to e^+ \pi^0) > 2.4 \cdot 10^{34}$ years~\cite{Super-Kamiokande:2020wjk}, excludes values of the GUT scale below $M_{\rm GUT} \sim 1.3 \cdot 10^{16}$ GeV, while the projected Hyper-Kamiokande sensitivity at 90\% C.L. after 20 years of runtime, $\tau(p \to e^+ \pi^0) > 1.2 \cdot 10^{35}$ years~\cite{Bhattiprolu:2022xhm}, would push this limit on the unification scale in our model to $M_{\rm GUT} \sim 2 \cdot 10^{16}$ GeV. Therefore, our model will be probed in the next round of proton decay searches, although large regions of the parameter space predict a long proton lifetime, well beyond any foreseen experiment.

\section{Conclusions \label{sec:conclusions}}
In this paper we have discussed 
$SU(5)^{3}$ with cyclic symmetry as a possible GUT. 
The basic idea of such a tri-unification is that there is
a separate $SU(5)$ for each fermion family, with the light Higgs doublet(s) arising from the third family $SU(5)$, providing a basis for charged fermion mass hierarchies. We have set out a general framework in which a class of such models which have been proposed in the literature, including $U(1)_Y^3$, $SU(2)_L^3$ and other related models, may have an ultraviolet completion in terms of $SU(5)^{3}$ tri-unification.

The main analysis in the paper was concerned with a
particular embedding of the tri-hypercharge model $U(1)_Y^3$ into $SU(5)^{3}$
with cyclic symmetry. 
We showed that a rather minimal example can account for all the quark and lepton (including neutrino) masses and mixing parameters. This same example can also satisfy the 
constraints of gauge coupling unification into the cyclic $SU(5)^{3}$ gauge group, by assuming minimal multiplet splitting, together with a set of relatively light colour octet scalars. The approximate conservation of the cyclic symmetry at low energies is also crucial to achieve gauge unification. The heavy messengers required to generate the flavour structure also modify the RGE in the desired way, highlighting the minimality of the framework.

Finally, we have also studied proton decay in this example, and presented the predictions of the proton lifetime in the dominant $e^+\pi^0$ channel. The results depend on the scale at which the three SM gauge groups break down into their diagonal non-Abelian subgroup together with tri-hypercharge, which is a free parameter in this model, enabling the proton lifetime to escape the existing Super-Kamiokande bound, but be possibly observable at Hyper-Kamiokande. In this manner, the signals on proton decay may allow to test the model at high scales, while low energy signals associated with tri-hypercharge enable the model to be tested by collider and flavour experiments. We conclude that $SU(5)^{3}$ tri-unification reconciles the idea of gauge non-universality with the idea of gauge coupling unification, opening up the possibility to build consistent non-universal descriptions of Nature that are valid all the way up to the scale of grand unification.

\section*{Acknowledgements}
MFN and AV are grateful to Renato Fonseca for many enlightening discussions on group theory, gauge unification and embeddings of SM fermions in GUTs.
SFK acknowledges the STFC Consolidated Grant ST/L000296/1 and the European Union's Horizon 2020 Research and Innovation programme under Marie Sklodowska-Curie grant agreement HIDDeN European ITN project (H2020-MSCA-ITN-2019//860881-HIDDeN). MFN is supported by the STFC under grant ST/X000605/1. AV acknowledges financial support from the Spanish grants PID2020-113775GB-I00 (AEI/10.13039/ 501100011033) and CIPROM/2021/054 (Generalitat Valenciana), as well as from MINECO through the Ram\'on y Cajal contract RYC2018-025795-I.

\appendix

\section{Energy regimes, symmetries and particle content}
\label{sec:app}

We describe the symmetries and particle content of our model at each energy regime between the GUT and electroweak scales.

\subsection*{Regime 1: $\boldsymbol{SU(5)^3}$ breaking scale $\boldsymbol{\to \left( SU(3) \times SU(2) \times U(1) \right)^3}$ breaking scale}

{
\begin{table}[t]
  \centering
  \begin{tabular}{cccccccccc}
      \toprule
      \textbf{Field} & $\boldsymbol{SU(3)_1}$ & $\boldsymbol{SU(2)_1}$ & $\boldsymbol{U(1)_1}$ & $\boldsymbol{SU(3)_2}$ & $\boldsymbol{SU(2)_2}$ & $\boldsymbol{U(1)_2}$ & $\boldsymbol{SU(3)_3}$ & $\boldsymbol{SU(2)_3}$ & $\boldsymbol{U(1)_3}$ \\
      \midrule
      $q_1$ &  \three  &  \two    &  $\frac{1}{6}$ & \one & \one & 0 & \one & \one & 0 \\
      $u_1^c$ &  \threeS   &  \one  & $-\frac{2}{3}$ & \one & \one & 0 & \one & \one & 0 \\
      $d_1^c$ &  \threeS   &  \one  & $\frac{1}{3}$ & \one & \one & 0 & \one & \one & 0 \\
      $\ell_1$ &  \one    &  \two    &  $-\frac{1}{2}$  & \one & \one & 0 & \one & \one & 0 \\
      $e_1^c$ &  \one    &  \one    &  1   & \one & \one & 0 & \one & \one & 0 \\
      $q_2$ &  \one & \one & 0 & \three  &  \two    &  $\frac{1}{6}$ & \one & \one & 0 \\
      $u_2^c$ &  \one & \one & 0 & \threeS   &  \one  & $-\frac{2}{3}$ & \one & \one & 0 \\
      $d_2^c$ &  \one & \one & 0 & \threeS   &  \one  & $\frac{1}{3}$ & \one & \one & 0 \\
      $\ell_2$ &  \one & \one & 0 & \one    &  \two    &  $-\frac{1}{2}$  & \one & \one & 0 \\
      $e_2^c$ &  \one & \one & 0 & \one    &  \one    &  1   & \one & \one & 0 \\
      $q_3$ & \one & \one & 0 & \one & \one & 0 &  \three  &  \two    &  $\frac{1}{6}$ \\
      $u_3^c$ & \one & \one & 0 & \one & \one & 0 &  \threeS   &  \one  & $-\frac{2}{3}$ \\
      $d_3^c$ & \one & \one & 0 & \one & \one & 0 &  \threeS   &  \one  & $\frac{1}{3}$ \\
      $\ell_3$ & \one & \one & 0 & \one & \one & 0 &  \one    &  \two    &  $-\frac{1}{2}$  \\
      $e_3^c$ & \one & \one & 0 & \one & \one & 0 &  \one    &  \one    &  1   \\
      \midrule
      $\xi_0$ & \one & \one & 0 & \one & \one & 0 &  \one    &  \one  &  0   \\
      \rowcolor{yellow!10} $\xi_{12}$ & \one & \one & $\frac{1}{2}$ & \one & \one & $-\frac{1}{2}$ &  \one    &  \one  &  0   \\
      \rowcolor{yellow!10} $\xi_{13}$ & \one & \one & $\frac{1}{2}$ & \one & \one & 0 &  \one    &  \one  &  $-\frac{1}{2}$   \\
       \rowcolor{yellow!10} $\xi_{23}$ & \one & \one & 0 & \one & \one & $\frac{1}{2}$ &  \one    &  \one  &  $-\frac{1}{2}$   \\
      \rowcolor{yellow!10} $Q_1$ &  \three  &  \two    &  $\frac{1}{6}$ & \one & \one & 0 & \one & \one & 0 \\
      \rowcolor{yellow!10} $Q_2$ &  \one & \one & 0 & \three  &  \two    &  $\frac{1}{6}$ & \one & \one & 0 \\
      \rowcolor{yellow!10} $Q_3$ &  \one & \one & 0 & \one & \one & 0 & \three  &  \two    &  $\frac{1}{6}$ \\
      \rowcolor{yellow!10} $N_{\rm atm}$ &  \one  &  \one    &  0 & \one & \one & $\frac{2}{3}$ & \one & \one & $-\frac{2}{3}$ \\
      \rowcolor{yellow!10} $N_{\rm sol}$ &  \one   &  \one  & $\frac{2}{3}$ & \one & \one & 0 & \one & \one & $-\frac{2}{3}$ \\
      \rowcolor{yellow!10} $N_{\rm cyclic}$ &  \one   &  \one  & $\frac{2}{3}$ & \one & \one & $-\frac{2}{3}$ & \one & \one & 0 \\
      \midrule
      $\Theta_1$    &  \eight   &  \one    &  0  &  \one   &  \one    &  0 &  \one   &  \one    &  0  \\
      $\Theta_2$    &  \one   &  \one    &  0  &  \eight   &  \one    &  0 &  \one   &  \one    &  0  \\
      $\Theta_3$    &  \one   &  \one    &  0  &  \one   &  \one    &  0 &  \eight   &  \one    &  0  \\
      $\Delta_1$    &  \one   &  \three    &  0  &  \one   &  \one    &  0 &  \one   &  \one    &  0  \\
      $\Delta_2$    &  \one   &  \one    &  0  &  \one   &  \three    &  0 &  \one   &  \one    &  0  \\
      $\Delta_3$    &  \one   &  \one    &  0  &  \one   &  \one    &  0 &  \one   &  \three    &  0  \\
      $H^u_1$    &  \one   &  \two    &  $\frac{1}{2}$  &  \one   &  \one    &  0 &  \one   &  \one    &  0  \\
      $H^d_1$    &  \one   &  \two    &  $-\frac{1}{2}$  &  \one   &  \one    &  0 &  \one   &  \one    &  0  \\
      $H^u_2$    &  \one   &  \one    &  0  &  \one   &  \two    &  $\frac{1}{2}$ &  \one   &  \one    &  0  \\
      $H^d_2$    &  \one   &  \one    &  0  &  \one   &  \two    &  $-\frac{1}{2}$ &  \one   &  \one    &  0  \\
      $H^u_3$    &  \one   &  \one    &  0  &  \one   &  \one    &  0 &  \one   &  \two    &  $\frac{1}{2}$  \\
      $H^d_3$    &  \one   &  \one    &  0  &  \one   &  \one    &  0 &  \one   &  \two    &  $-\frac{1}{2}$  \\
      \midrule
      $\Phi_{\ell 12}$ & \one & \one & $\frac{1}{2}$ & \one & \one & $-\frac{1}{2}$ &  \one    &  \one  &  0   \\
      $\Phi_{\ell 13}$ & \one & \one & $\frac{1}{2}$ & \one & \one & 0 &  \one    &  \one  &  $-\frac{1}{2}$   \\
      $\Phi_{\ell 23}$ & \one & \one & 0 & \one & \one & $\frac{1}{2}$ &  \one    &  \one  &  $-\frac{1}{2}$   \\
      $\Phi_{q 12}$ & \one & \one & $-\frac{1}{6}$ & \one & \one & $\frac{1}{6}$ &  \one    &  \one  &  0   \\
      $\Phi_{q 13}$ & \one & \one & $-\frac{1}{6}$ & \one & \one & 0 &  \one    &  \one  &  $\frac{1}{6}$   \\
      $\Phi_{q 23}$ & \one & \one & 0 & \one & \one & $-\frac{1}{6}$ &  \one    &  \one  &  $\frac{1}{6}$   \\
      $\Phi_{u 12}$ & \one & \one & $-\frac{2}{3}$ & \one & \one & $\frac{2}{3}$ &  \one    &  \one  &  0   \\
      $\Phi_{u 13}$ & \one & \one & $-\frac{2}{3}$ & \one & \one & 0 &  \one    &  \one  &  $\frac{2}{3}$   \\
      $\Phi_{u 23}$ & \one & \one & 0 & \one & \one & $-\frac{2}{3}$ &  \one    &  \one  &  $\frac{2}{3}$   \\
      \bottomrule
  \end{tabular}
  \caption{
    Fermion and scalar representations under $\left( SU(3) \times SU(2) \times U(1) \right)^3$ in energy regime 1. Fermions highlighted in yellow belong to a vector-like pair and thus have a conjugate representation not shown in this table.
    \label{tab:ParticleContent2}
    }
\end{table}
}

As a result of \555 breaking, each of the fermion representations $F_i$ and $T_i$ becomes charged under an $SU(3) \times SU(2) \times U(1)$ factor. Regarding the rest of the
fields, most get masses at the $M_{\rm GUT} \sim v_{\mathrm{GUT}}$ unification scale and decouple. We will assume that only those explicitly required at low energies remain light. For instance, out of all the components of the $\Omega_{ij}$ scalars, only the $\Theta_i$ and $\Delta_i$ states, belonging to the adjoint representations of $SU(3)_i$ and $SU(2)_i$, respectively, remain in the particle spectrum. Similarly, only some SM singlets in the $\Phi_i$ scalar fields are assumed to be present at this energy scale. For instance, this is the case of $\Phi_{\ell 23}$, contained in
$\Phi_{23}^{(5)}$, a (\one,\five,\fiveS) representation of \555, as shown in Table~\ref{tab:hyperons1}. These representations eventually become the tri-hypercharge hyperons at lower energies. Similarly, the $Q_i$ vector-like quarks in the $\chi_i$ and $\overline{\chi}_i$ multiplets are also assumed to be present at this energy scale. The full fermion and scalar particle content of the model in this energy regime is shown in Table~\ref{tab:ParticleContent2}.

\subsection*{Regime 2: $\boldsymbol{\left( SU(3) \times SU(2) \times U(1) \right)^3}$ breaking scale $\boldsymbol{\to \xi}$ scale}

{
\begin{table}[tb]
  \centering
  \begin{tabular}{cccccc}
      \toprule
      \textbf{Field} & $\boldsymbol{SU(3)_c}$ & $\boldsymbol{SU(2)_L}$ & $\boldsymbol{U(1)_{Y_1}}$ & $\boldsymbol{U(1)_{Y_2}}$ & $\boldsymbol{U(1)_{Y_3}}$ \\
      \midrule
      $q_1$ &  \three  &  \two    &  $\frac{1}{6}$ & 0 & 0 \\
      $u_1^c$ &  \threeS   &  \one  & $-\frac{2}{3}$ & 0 & 0 \\
      $d_1^c$ &  \threeS   &  \one  & $\frac{1}{3}$ & 0 & 0 \\
      $\ell_1$ &  \one    &  \two    &  $-\frac{1}{2}$  & 0 & 0 \\
      $e_1^c$ &  \one    &  \one    &  1  & 0 & 0 \\
      $q_2$ &  \three  &  \two  & 0 & $\frac{1}{6}$ & 0 \\
      $u_2^c$ &  \threeS   &  \one & 0 & $-\frac{2}{3}$ & 0 \\
      $d_2^c$ &  \threeS   &  \one  & 0 & $\frac{1}{3}$ & 0 \\
      $\ell_2$ &  \one    &  \two    & 0 & $-\frac{1}{2}$  & 0 \\
      $e_2^c$ &  \one    &  \one    &  0 & 1  & 0 \\
      $q_3$ & \three  &  \two    & 0 & 0 & $\frac{1}{6}$ \\
      $u_3^c$ & \threeS   &  \one  & 0 & 0 & $-\frac{2}{3}$ \\
      $d_3^c$ & \threeS   &  \one  & 0 & 0 & $\frac{1}{3}$ \\
      $\ell_3$ & \one    &  \two    &  0 & 0 & $-\frac{1}{2}$  \\
      $e_3^c$ & \one    &  \one    &  0 & 0 & 1   \\
      \midrule
      $\xi_0$ & \one & \one & 0 & 0 & 0   \\
      \rowcolor{yellow!10} $\xi_{12}$ & \one & \one & $\frac{1}{2}$ & $-\frac{1}{2}$ &  0   \\
      \rowcolor{yellow!10} $\xi_{13}$ & \one & \one & $\frac{1}{2}$ & 0 &  $-\frac{1}{2}$   \\
      \rowcolor{yellow!10} $\xi_{23}$ & \one & \one & 0 & $\frac{1}{2}$ & $-\frac{1}{2}$   \\
      \rowcolor{yellow!10} $Q_1$ &  \three  &  \two    &  $\frac{1}{6}$ & 0 & 0 \\
      \rowcolor{yellow!10} $Q_2$ &  \three  &  \two    &  0 & $\frac{1}{6}$ & 0 \\
      \rowcolor{yellow!10} $Q_3$ &  \three  &  \two    &  0 & 0 & $\frac{1}{6}$ \\
      \rowcolor{yellow!10} $N_{\rm atm}$ &  \one  &  \one    &  0 & $\frac{2}{3}$ & $-\frac{2}{3}$ \\
      \rowcolor{yellow!10} $N_{\rm sol}$ &  \one   &  \one  & $\frac{2}{3}$ & 0 & $-\frac{2}{3}$ \\
      \midrule
      $\Theta_1$    &  \eight   &  \one    &  0  &  0 &  0  \\
      $\Theta_2$    &  \eight   &  \one    &  0  &  0 &  0  \\
      $\Theta_3$    &  \eight   &  \one    &  0  &  0 &  0  \\
      $H^u_1$    &  \one   &  \two    &  $\frac{1}{2}$  &  0 &  0  \\
      $H^d_1$    &  \one   &  \two    &  $-\frac{1}{2}$  &  0 &  0  \\
      $H^u_2$    &  \one   &  \two    &  0  & $\frac{1}{2}$ &  0  \\
      $H^d_2$    &  \one   &  \two    &  0  & $-\frac{1}{2}$  &  0  \\
      $H^u_3$    &  \one   &  \two    &  0  &  0  &  $\frac{1}{2}$  \\
      $H^d_3$    &  \one   &  \two    &  0  &  0  &  $-\frac{1}{2}$ \\
      \midrule
      $\phi_{\ell 12}$ & \one & \one & $\frac{1}{2}$ & $-\frac{1}{2}$ & 0  \\
      $\phi_{\ell 13}$ & \one & \one & $\frac{1}{2}$ & 0 & $-\frac{1}{2}$   \\
      $\phi_{\ell 23}$ & \one & \one & 0 & $\frac{1}{2}$ & $-\frac{1}{2}$   \\
      $\phi_{q 12}$ & \one & \one & $-\frac{1}{6}$ & $\frac{1}{6}$ & 0   \\
      $\phi_{q 13}$ & \one & \one & $-\frac{1}{6}$ & 0 & $\frac{1}{6}$   \\
      $\phi_{q 23}$ & \one & \one & 0 & $-\frac{1}{6}$ & $\frac{1}{6}$   \\
      $\phi_{u 12}$ & \one & \one & $-\frac{2}{3}$ & $\frac{2}{3}$ & 0   \\
      $\phi_{u 13}$ & \one & \one & $-\frac{2}{3}$ & 0 & $\frac{2}{3}$   \\
      $\phi_{u 23}$ & \one & \one & 0 & $-\frac{2}{3}$ & $\frac{2}{3}$   \\      
      \bottomrule
  \end{tabular}
  \caption{
    Fermion and scalar representations under $SU(3)_c \times SU(2)_L \times U(1)_{Y_1} \times U(1)_{Y_2} \times U(1)_{Y_3}$ in energy regimes 2, 3, 4, 5 and 6. Some states in this table get decoupled at intermediate scales and are not present at all energy regimes, see text for details. Fermions highlighted in yellow belong to a vector-like pair and thus have a conjugate representation not shown in this table. 
    \label{tab:ParticleContent3}
    }
\end{table}
}

The $\left( SU(3) \times SU(2) \times U(1) \right)^3$ gauge symmetry gets broken by the non-zero VEVs of the $\Theta_i$ and $\Delta_i$ scalars. The $\Theta_i$ octets break $SU(3)_1 \times SU(3)_2 \times SU(3)_3 \to SU(3)_{1+2+3} \equiv SU(3)_c$, while the $\Delta_i$ triplets play an analogous role for the $SU(2)$ factors. We assume these two breakings to take place simultaneously at $v_{\rm{SM}^3} = \langle \Theta_i \rangle = \langle \Delta_i \rangle$, slightly below the GUT scale. As a result of this, the remnant symmetry is the tri-hypercharge group~\cite{FernandezNavarro:2023rhv}, $SU(3)_c \times SU(2)_L \times U(1)_{Y_1} \times U(1)_{Y_2} \times
U(1)_{Y_3}$:
\begin{equation}
  \left( SU(3) \times SU(2) \times U(1) \right)^3 \quad \xrightarrow{\langle \Theta_i \rangle,\langle \Delta_i \rangle} \quad SU(3)_c \times SU(2)_L \times U(1)_{Y_1} \times U(1)_{Y_2} \times U(1)_{Y_3}
\end{equation}
The gauge couplings above ($g_{s_i}$ and $g_{L_i}$, with $i=1,2,3$)
and below ($g_s$ and $g_L$) the breaking scale verify the matching
relations
\begin{align}
  \frac{g_{s_1} \, g_{s_2} \, g_{s_3}}{\sqrt{g_{s_1}^2 g_{s_2}^2 + g_{s_1}^2 g_{s_3}^2 + g_{s_2}^2 g_{s_3}^2}} &= g_s \, , \\
  \frac{g_{L_1} \, g_{L_2} \, g_{L_3}}{\sqrt{g_{L_1}^2 g_{L_2}^2 + g_{L_1}^2 g_{L_3}^2 + g_{L_2}^2 g_{L_3}^2}} &= g_L \, ,
\end{align}
which are equivalent to
\begin{align}
  \alpha_{s_1}^{-1} + \alpha_{s_2}^{-1} + \alpha_{s_3}^{-1} &= \alpha_s^{-1} \, , \\
  \alpha_{L_1}^{-1} + \alpha_{L_2}^{-1} + \alpha_{L_3}^{-1} &= \alpha_L^{-1} \, , 
\end{align}
with $\alpha_i^{-1} = 4 \pi/g_i^2$.

The main difference with respect to the original tri-hypercharge model~\cite{FernandezNavarro:2023rhv} is that a complete ultraviolet completion for the generation of the flavour structure is provided in our setup. As already explained, we achieve this with the hyperons and vector-like fermions present in the particle spectrum, which originate from \555 representations. We assume $N_{\rm cyclic}$ as well as the conjugate representation $\overline{N}_{\rm cyclic}$ to be decoupled at this energy scale. Similarly, the $\Delta_i$ triplets are also assumed to get masses of the order of the SM$^3$ breaking scale and decouple. The resulting fermion and scalar particle content of the model is shown in Table~\ref{tab:ParticleContent3}.

\subsection*{Regime 3: $\boldsymbol{\xi}$ scale $\boldsymbol{\to H_1}$ scale}

The next energy threshold is given by the $\xi$ singlets, responsible for the flavour structure of the neutrino sector, with masses $M_\xi \sim 10^{10}$ GeV. At this scale, the $\xi_0$ as well as the $\xi_{12}$, $\xi_{13}$, $\xi_{23}$ and their conjugate representations are integrated out and no longer contribute to the running of the gauge couplings. The gauge symmetry does not change and stays the same as in the previous energy regime. The resulting particle spectrum is that of Table~\ref{tab:ParticleContent3} removing the $\xi$ singlet fermions.

\subsection*{Regime 4: $\boldsymbol{H_1}$ scale $\boldsymbol{\to H_2}$ scale}

At energies of the order of $M_{H_1^{u,d}} \sim 10^4$ TeV, the $H_1^{u,d}$ scalar doublets decouple from the particle spectrum of the model. Again, the gauge symmetry does not change. The particle spectrum at this stage is that shown on Table~\ref{tab:ParticleContent3} removing the $\xi$ singlet fermions and the $H^{u,d}_1$ scalar doublets.

\subsection*{Regime 5: $\boldsymbol{H_2}$ scale $\boldsymbol{\to Q,\,\Theta}$ scale}

At energies of the order of $M_{H_2^{u,d}} \sim 100$ TeV, the $H_2^{u,d}$ scalar doublets decouple from the particle spectrum of the model. As in the previous two energy thresholds, the gauge symmetry remains the same. The particle spectrum at this stage is that shown on Table~\ref{tab:ParticleContent3} removing the $\xi$ singlet fermions and the $H^{u,d}_{1,2}$ scalar doublets.

\subsection*{Regime 6: $\boldsymbol{Q,\,\Theta}$ scale $\boldsymbol{\to SU(3)_c \times SU(2)_L \times U(1)_{Y_1} \times U(1)_{Y_2} \times U(1)_{Y_3}}$ breaking scale}

{
\begin{table}[tb]
  \centering
  \begin{tabular}{ccccc}
      \toprule
      \textbf{Field} & $\boldsymbol{SU(3)_c}$ & $\boldsymbol{SU(2)_L}$ & $\boldsymbol{U(1)_{Y_{12}}}$ & $\boldsymbol{U(1)_{Y_3}}$ \\
      \midrule
      $q_1$ &  \three  &  \two    &  $\frac{1}{6}$ & 0 \\
      $u_1^c$ &  \threeS   &  \one  & $-\frac{2}{3}$ & 0 \\
      $d_1^c$ &  \threeS   &  \one  & $\frac{1}{3}$ & 0 \\
      $\ell_1$ &  \one    &  \two    &  $-\frac{1}{2}$  & 0 \\
      $e_1^c$ &  \one    &  \one    &  1  & 0 \\
      $q_2$ &  \three  &  \two  & $\frac{1}{6}$ & 0 \\
      $u_2^c$ &  \threeS   &  \one & $-\frac{2}{3}$ & 0 \\
      $d_2^c$ &  \threeS   &  \one  & $\frac{1}{3}$ & 0 \\
      $\ell_2$ &  \one    &  \two    & $-\frac{1}{2}$  & 0 \\
      $e_2^c$ &  \one    &  \one    &  1  & 0 \\
      $q_3$ & \three  &  \two    & 0 & $\frac{1}{6}$ \\
      $u_3^c$ & \threeS   &  \one  & 0 & $-\frac{2}{3}$ \\
      $d_3^c$ & \threeS   &  \one  & 0 & $\frac{1}{3}$ \\
      $\ell_3$ & \one    &  \two    &  0 & $-\frac{1}{2}$  \\
      $e_3^c$ & \one    &  \one    &  0 & 1   \\
      \midrule
      \rowcolor{yellow!10} $N_{\rm atm}$ &  \one  &  \one    &  $\frac{2}{3}$ & $-\frac{2}{3}$ \\
      \rowcolor{yellow!10} $N_{\rm sol}$ &  \one   &  \one  & $\frac{2}{3}$ & $-\frac{2}{3}$ \\
      \midrule
      $H^u_3$    &  \one   &  \two    &  0  &  $\frac{1}{2}$  \\
      $H^d_3$    &  \one   &  \two    &  0  &  $-\frac{1}{2}$ \\
      \midrule
      $\phi_{\ell 13}$ & \one & \one & $\frac{1}{2}$ & $-\frac{1}{2}$   \\
      $\phi_{\ell 23}$ & \one & \one & $\frac{1}{2}$ & $-\frac{1}{2}$   \\
      $\phi_{q 13}$ & \one & \one & $-\frac{1}{6}$ & $\frac{1}{6}$   \\
      $\phi_{q 23}$ & \one & \one & $-\frac{1}{6}$ & $\frac{1}{6}$   \\
      $\phi_{u 13}$ & \one & \one & $-\frac{2}{3}$ & $\frac{2}{3}$   \\
      $\phi_{u 23}$ & \one & \one & $-\frac{2}{3}$ & $\frac{2}{3}$   \\      
      \bottomrule
  \end{tabular}
  \caption{
    Fermion and scalar representations under $SU(3)_c \times SU(2)_L \times U(1)_{Y_{12}} \times U(1)_{Y_3}$ in energy regime 7. Fermions highlighted in yellow belong to a vector-like pair and thus have a conjugate representation not shown in this table.
    \label{tab:ParticleContent4}
    }
\end{table}
}

At $M_Q \lesssim M_{H_2^{u,d}}$, the $Q_i$ vector-like quarks and the $\Theta_{i}$ colour octets decouple from the particle spectrum of the model. As in the previous two energy thresholds, the gauge symmetry is not altered. The particle spectrum at this stage is that shown on Table~\ref{tab:ParticleContent3} removing the $\xi$ singlet fermions, the $H^{u,d}_{1,2}$ scalar doublets, the $Q_i$ vector-like quarks and the $\Theta_{i}$ colour octets.

Hyperons are responsible for the breaking of the tri-hypercharge symmetry. In a first hypercharge breaking step, $U(1)_{Y_1} \times U(1)_{Y_2} \times U(1)_{Y_3}$ gets broken to $U(1)_{Y_{12}} \times U(1)_{Y_3}$, where $Y_{12} = Y_1 + Y_2$, by the non-zero VEV of the
$\phi_{q 12}$ hyperon, $v_{12} = \langle \phi_{q 12} \rangle \sim 50$ TeV:
\begin{equation}
  SU(3)_c \times SU(2)_L \times U(1)_{Y_1} \times U(1)_{Y_2} \times U(1)_{Y_3} \quad \xrightarrow{\langle \phi_{q 12} \rangle} \quad SU(3)_c \times SU(2)_L \times U(1)_{Y_{12}} \times U(1)_{Y_3}
\end{equation}
The gauge couplings above ($g_{Y_1}$ and $g_{Y_2}$) and below ($g_{Y_{12}}$) the breaking scale verify the matching relation
\begin{equation}
  \frac{g_{Y_1} \, g_{Y_2}}{\sqrt{g_{Y_1}^2 + g_{Y_2}^2}} = g_{Y_{12}} \, ,
\end{equation}
which is equivalent to
\begin{equation}
  \alpha_{Y_1}^{-1} + \alpha_{Y_2}^{-1} = \alpha_{Y_{12}}^{-1} \, .
\end{equation}
The ``12 hyperons'' $\phi_{\ell 12}$, $\phi_{q 12}$ and $\phi_{u 12}$ get masses of the order of $\langle \phi_{q 12} \rangle$ and decouple at this stage. We also assume the $\Theta_i$ colour octets to be integrated out at the tri-hypercharge breaking scale. The resulting fermion and scalar particle content is shown in Table~\ref{tab:ParticleContent4}.

\subsection*{Regime 7: $\boldsymbol{SU(3)_c \times SU(2)_L \times U(1)_{Y_1} \times U(1)_{Y_2} \times U(1)_{Y_3}}$ breaking scale\\ \hspace*{2cm} $\boldsymbol{\to SU(3)_c \times SU(2)_L \times U(1)_{Y_{12}} \times U(1)_{Y_3}}$ breaking scale}

The $SU(3)_c \times SU(2)_L \times U(1)_{Y_{12}} \times U(1)_{Y_3}$ gauge symmetry also gets broken by hyperon VEVs, leaving as a remnant the conventional SM gauge symmetry with $Y = Y_{12} + Y_3 = Y_1 + Y_2 + Y_3$. In this case, the hyperons responsible for the breaking are $\phi_{\ell 13}$, $\phi_{\ell 23}$, $\phi_{q 13}$ and $\phi_{q 23}$, which get VEVs of the order of $v_{23} \sim 5$ TeV:
\begin{equation}
  SU(3)_c \times SU(2)_L \times U(1)_{Y_{12}} \times U(1)_{Y_3} \quad \xrightarrow{\langle \phi_{\ell 13, 23} \rangle,\langle \phi_{q 13, 23} \rangle} \quad SU(3)_c \times SU(2)_L \times U(1)_Y
\end{equation}
The gauge couplings above ($g_{Y_{12}}$ and $g_{Y_3}$) and below ($g_Y$) the breaking scale verify the matching relation
\begin{equation}
  \frac{g_{Y_{12}} \, g_{Y_3}}{\sqrt{g_{Y_{12}}^2 + g_{Y_3}^2}} = g_Y \, ,
\end{equation}
which is equivalent to
\begin{equation}
  \alpha_{Y_{12}}^{-1} + \alpha_{Y_3}^{-1} = \alpha_Y^{-1} \, .
\end{equation}
All the remaining hyperons as well as the neutrino mass messengers $N_{\rm atm}$ and $N_{\rm sol}$ (as well as their conjugate representations) decouple at this stage. The resulting particle spectrum is that of a two Higgs doublet model, with universal charges for all fermions.

\subsection*{Regime 8: $\boldsymbol{SU(3)_c \times SU(2)_L \times U(1)_{Y_{12}} \times U(1)_{Y_3}}$ breaking scale\\ \hspace*{2cm} $\boldsymbol{\to SU(3)_c \times SU(2)_L \times U(1)_Y}$ breaking scale}

Finally, at the scale $v_{\mathrm{SM}}$, the electroweak symmetry gets broken in the usual way, by the VEVs of the $H_3^{u,d}$ scalar doublets:
\begin{equation}
  SU(3)_c \times SU(2)_L \times U(1)_Y \quad \xrightarrow{\langle H_3^{u,d} \rangle} \quad SU(3)_c \times U(1)_{\rm em}
\end{equation}

\section{Hyperons from \texorpdfstring{$\boldsymbol{SU(5)^{3}}$}{SU(5)^{3}}} \label{sec:Hyperons}

Tables~\ref{tab:hyperons1} and \ref{tab:hyperons2} list all possible hyperon embeddings in $SU(5)^{3}$ representations with dimension up to $\mathbf{45}$. These tables have been obtained with the help of~\texttt{GroupMath}~\cite{Fonseca:2020vke}.

{
\renewcommand{\arraystretch}{1.4}
\begin{table}[tb]
  \footnotesize
  \centering
  \begin{tabular}{|rrr|c|}
      \hline
      \multicolumn{3}{|c|}{\textbf{Hyperon}} & \textbf{$\boldsymbol{SU(5)^{3}}$ representations} \\
      \hline
\rowcolor{blue!10} 0 & $-\frac{1}{3}$ & $\frac{1}{3}$ & (\one,\five,\fiveS),(\one,\five,\ffiveS),(\one,\ffive,\fiveS),(\one,\ffive,\ffiveS),(\24,\five,\fiveS),(\24,\five,\ffiveS),(\24,\ffive,\fiveS),(\24,\ffive,\ffiveS) \\
  0 & $\frac{1}{2}$ & $-\frac{1}{2}$ & (\one,\five,\fiveS),(\one,\five,\ffiveS),(\one,\ffive,\fiveS),(\one,\ffive,\ffiveS),(\24,\five,\fiveS),(\24,\five,\ffiveS),(\24,\ffive,\fiveS),(\24,\ffive,\ffiveS) \\
  \rowcolor{blue!10} & & & (\one,\ten,\tenS),(\one,\ten,\fortyS),(\one,\fifteen,\fifteenS),(\one,\tfive,\tfiveS),(\one,\tfive,\fortyS),(\one,\forty,\tenS),(\one,\forty,\tfiveS),(\one,\forty,\fortyS), \\
  \rowcolor{blue!10} & & & (\24,\ten,\tenS),(\24,\ten,\fifteenS),(\24,\ten,\tfiveS),(\24,\ten,\fortyS),(\24,\fifteen,\tenS),(\24,\fifteen,\fifteenS),(\24,\fifteen,\fortyS), \\
  \rowcolor{blue!10} \multirow{-3}{*}{0} & \multirow{-3}{*}{$-\frac{2}{3}$} & \multirow{-3}{*}{$\frac{2}{3}$} & (\24,\tfive,\tenS),(\24,\tfive,\tfiveS),(\24,\tfive,\fortyS),(\24,\forty,\tenS),(\24,\forty,\fifteenS),(\24,\forty,\tfiveS),(\24,\forty,\fortyS) \\
  & & & (\one,\ten,\tenS),(\one,\ten,\fifteenS),(\one,\ten,\fortyS),(\one,\fifteen,\tenS),(\one,\fifteen,\fifteenS),(\one,\fifteen,\fortyS),(\one,\tfive,\tfiveS), \\
  & & & (\one,\tfive,\fortyS),(\one,\forty,\tenS),(\one,\forty,\fifteenS),(\one,\forty,\tfiveS),(\one,\forty,\fortyS),(\24,\ten,\tenS),(\24,\ten,\fifteenS), \\
  & & & (\24,\ten,\tfiveS),(\24,\ten,\fortyS),(\24,\fifteen,\tenS),(\24,\fifteen,\fifteenS),(\24,\fifteen,\tfiveS),(\24,\fifteen,\fortyS),(\24,\tfive,\tenS), \\
  \multirow{-4}{*}{0} & \multirow{-4}{*}{$\frac{1}{6}$} & \multirow{-4}{*}{$-\frac{1}{6}$} & (\24,\tfive,\fifteenS),(\24,\tfive,\tfiveS),(\24,\tfive,\fortyS),(\24,\forty,\tenS),(\24,\forty,\fifteenS),(\24,\forty,\tfiveS),(\24,\forty,\fortyS) \\
  \rowcolor{blue!10} & & & (\one,\ten,\tenS),(\one,\fifteen,\fifteenS),(\one,\tfive,\tfiveS),(\one,\forty,\fortyS),(\24,\ten,\tenS),(\24,\ten,\fifteenS),(\24,\ten,\fortyS), \\
  \rowcolor{blue!10} \multirow{-2}{*}{0} & \multirow{-2}{*}{$1$} & \multirow{-2}{*}{$-1$} & (\24,\fifteen,\tenS),(\24,\fifteen,\fifteenS),(\24,\tfive,\tfiveS),(\24,\tfive,\fortyS),(\24,\forty,\tenS),(\24,\forty,\tfiveS),(\24,\forty,\fortyS) \\
  0 & $\frac{5}{6}$ & $-\frac{5}{6}$ & (\one,\24,\24), (\24,\24,\24) \\
  \rowcolor{blue!10} 0 & $-\frac{3}{2}$ & $\frac{3}{2}$ & (\one,\tfive,\tfiveS),(\one,\forty,\fortyS),(\24,\tfive,\tfiveS),(\24,\tfive,\fortyS),(\24,\forty,\tfiveS),(\24,\forty,\fortyS) \\
  0 & $\frac{4}{3}$ & $-\frac{4}{3}$ & (\one,\ffive,\ffiveS), (\24,\ffive,\ffiveS) \\
  \rowcolor{blue!10} 0 & $-\frac{7}{6}$ & $\frac{7}{6}$ & (\one,\ffive,\ffiveS), (\24,\ffive,\ffiveS) \\
      \hline
  \end{tabular}
  \caption{
  \small Hyperons charged under two individual hypercharge groups and their \555 origin. All \555 representations that involve up to $\ffive$ and $\ffiveS$ of $\GGM$ are included. Other hyperons can be obtained by reordering the hypercharge values or by conjugating the \555 representations.
  \label{tab:hyperons1}
  }
\end{table}
}

{
\renewcommand{\arraystretch}{1.4}
\begin{table}[tb]
  \footnotesize
  \centering
  \begin{tabular}{|rrr|c|}
      \hline
      \multicolumn{3}{|c|}{\textbf{Hyperon}} & \textbf{$\boldsymbol{SU(5)^{3}}$ representations} \\
      \hline
  & & & (\five,\five,\tenS),(\five,\five,\fifteenS),(\five,\five,\fortyS),(\five,\ffive,\tenS),(\five,\ffive,\fifteenS),(\five,\ffive,\tfiveS),(\five,\ffive,\fortyS),(\ffive,\five,\tenS), \\
  \multirow{-2}{*}{$-\frac{1}{3}$} & \multirow{-2}{*}{$-\frac{1}{3}$} & \multirow{-2}{*}{$\frac{2}{3}$} & (\ffive,\five,\fifteenS),(\ffive,\five,\tfiveS),(\ffive,\five,\fortyS),(\ffive,\ffive,\tenS),(\ffive,\ffive,\fifteenS),(\ffive,\ffive,\tfiveS),(\ffive,\ffive,\fortyS) \\
  \rowcolor{blue!10} & & & (\five,\five,\tenS),(\five,\five,\fifteenS),(\five,\five,\fortyS),(\five,\ffive,\tenS),(\five,\ffive,\fifteenS),(\five,\ffive,\tfiveS),(\five,\ffive,\fortyS),(\ffive,\five,\tenS), \\
  \rowcolor{blue!10} \multirow{-2}{*}{$-\frac{1}{3}$} & \multirow{-2}{*}{$\frac{1}{2}$} & \multirow{-2}{*}{$-\frac{1}{6}$} & (\ffive,\five,\fifteenS),(\ffive,\five,\tfiveS),(\ffive,\five,\fortyS),(\ffive,\ffive,\tenS),(\ffive,\ffive,\fifteenS),(\ffive,\ffive,\tfiveS),(\ffive,\ffive,\fortyS) \\
  & & & (\five,\five,\tenS),(\five,\five,\fifteenS),(\five,\ffive,\tenS),(\five,\ffive,\fifteenS),(\five,\ffive,\fortyS),(\ffive,\five,\tenS), \\
  \multirow{-2}{*}{$\frac{1}{2}$} & \multirow{-2}{*}{$\frac{1}{2}$} & \multirow{-2}{*}{$-1$} & (\ffive,\five,\fifteenS),(\ffive,\five,\fortyS),(\ffive,\ffive,\tenS),(\ffive,\ffive,\fifteenS),(\ffive,\ffive,\tfiveS),(\ffive,\ffive,\fortyS) \\
  \rowcolor{blue!10} $-\frac{1}{3}$ & $-\frac{1}{2}$ & $\frac{5}{6}$ & (\five,\fiveS,\24),(\five,\ffiveS,\24),(\ffive,\fiveS,\24),(\ffive,\ffiveS,\24) \\
  & & & (\five,\ten,\ten),(\five,\ten,\forty),(\five,\fifteen,\tfive),(\five,\fifteen,\forty),(\five,\tfive,\fifteen),(\five,\forty,\ten),(\five,\forty,\fifteen), \\
  & & & (\five,\forty,\forty),(\ffive,\ten,\ten),(\ffive,\ten,\fifteen),(\ffive,\ten,\forty),(\ffive,\fifteen,\ten),(\ffive,\fifteen,\tfive),(\ffive,\fifteen,\forty), \\
  \multirow{-3}{*}{$-\frac{1}{3}$} & \multirow{-3}{*}{$-\frac{2}{3}$} & \multirow{-3}{*}{$1$} & (\ffive,\tfive,\ten),(\ffive,\tfive,\fifteen),(\ffive,\tfive,\forty),(\ffive,\forty,\ten),(\ffive,\forty,\fifteen),(\ffive,\forty,\forty) \\
  \rowcolor{blue!10} & & & (\five,\ten,\ten),(\five,\ten,\fifteen),(\five,\ten,\tfive),(\five,\ten,\forty),(\five,\fifteen,\ten),(\five,\fifteen,\fifteen),(\five,\fifteen,\tfive), \\
  \rowcolor{blue!10} & & & (\five,\fifteen,\forty),(\five,\tfive,\ten),(\five,\tfive,\fifteen),(\five,\tfive,\forty),(\five,\forty,\ten),(\five,\forty,\fifteen),(\five,\forty,\tfive), \\
  \rowcolor{blue!10} & & & (\five,\forty,\forty),(\ffive,\ten,\ten),(\ffive,\ten,\fifteen),(\ffive,\ten,\tfive),(\ffive,\ten,\forty),(\ffive,\fifteen,\ten), \\
  \rowcolor{blue!10} & & & (\ffive,\fifteen,\fifteen),(\ffive,\fifteen,\tfive),(\ffive,\fifteen,\forty),(\ffive,\tfive,\ten),(\ffive,\tfive,\fifteen),(\ffive,\tfive,\tfive), \\
  \rowcolor{blue!10} \multirow{-5}{*}{$-\frac{1}{3}$} & \multirow{-5}{*}{$\frac{1}{6}$} & \multirow{-5}{*}{$\frac{1}{6}$} & (\ffive,\tfive,\forty),(\ffive,\forty,\ten),(\ffive,\forty,\fifteen),(\ffive,\forty,\tfive),(\ffive,\forty,\forty) \\
  & & & (\five,\ten,\ten),(\five,\ten,\fifteen),(\five,\ten,\forty),(\five,\fifteen,\tfive),(\five,\fifteen,\forty),(\five,\tfive,\ten),(\five,\tfive,\fifteen), \\
  & & & (\five,\tfive,\forty),(\five,\forty,\ten),(\five,\forty,\fifteen),(\five,\forty,\forty),(\ffive,\ten,\ten),(\ffive,\ten,\fifteen),(\ffive,\ten,\tfive), \\
  & & & (\ffive,\ten,\forty),(\ffive,\fifteen,\ten),(\ffive,\fifteen,\fifteen),(\ffive,\fifteen,\tfive),(\ffive,\fifteen,\forty),(\ffive,\tfive,\ten),(\ffive,\tfive,\fifteen), \\
  \multirow{-4}{*}{$\frac{1}{2}$} & \multirow{-4}{*}{$-\frac{2}{3}$} & \multirow{-4}{*}{$\frac{1}{6}$} & (\ffive,\tfive,\tfive),(\ffive,\tfive,\forty),(\ffive,\forty,\ten),(\ffive,\forty,\fifteen),(\ffive,\forty,\tfive),(\ffive,\forty,\forty) \\
  \rowcolor{blue!10} $\frac{1}{2}$ & $1$ & $-\frac{3}{2}$ & (\five,\ten,\forty),(\five,\fifteen,\tfive),(\five,\fifteen,\forty),(\ffive,\ten,\forty),(\ffive,\fifteen,\tfive),(\ffive,\fifteen,\forty),(\ffive,\forty,\forty) \\
  $-\frac{1}{3}$ & $-1$ & $\frac{4}{3}$ & (\five,\tenS,\ffive),(\five,\fortyS,\ffive),(\ffive,\tenS,\ffive),(\ffive,\fifteenS,\ffive),(\ffive,\tfiveS,\ffive),(\ffive,\fortyS,\ffive) \\
  \rowcolor{blue!10} $\frac{1}{2}$ & $\frac{2}{3}$ & $-\frac{7}{6}$ & (\five,\tenS,\ffive),(\five,\tfiveS,\ffive),(\five,\fortyS,\ffive),(\ffive,\tenS,\ffive),(\ffive,\fifteenS,\ffive),(\ffive,\tfiveS,\ffive),(\ffive,\fortyS,\ffive) \\
  $-\frac{1}{3}$ & $-\frac{5}{6}$ & $\frac{7}{6}$ & (\five,\24,\ffiveS),(\ffive,\24,\ffiveS) \\
  \rowcolor{blue!10} $\frac{1}{2}$ & $\frac{5}{6}$ & $-\frac{4}{3}$ & (\five,\24,\ffiveS),(\ffive,\24,\ffiveS) \\
  $-\frac{1}{3}$ & $\frac{3}{2}$ & $-\frac{7}{6}$ & (\five,\fortyS,\ffive),(\ffive,\tfiveS,\ffive),(\ffive,\fortyS,\ffive) \\
  \rowcolor{blue!10} & & & (\ten,\ten,\ffive),(\ten,\fifteen,\ffive),(\ten,\forty,\ffive),(\fifteen,\ten,\ffive),(\fifteen,\forty,\ffive),(\tfive,\tfive,\ffive), \\
  \rowcolor{blue!10} \multirow{-2}{*}{$-\frac{2}{3}$} & \multirow{-2}{*}{$-\frac{2}{3}$} & \multirow{-2}{*}{$\frac{4}{3}$} & (\tfive,\forty,\ffive),(\forty,\ten,\ffive),(\forty,\fifteen,\ffive),(\forty,\tfive,\ffive),(\forty,\forty,\ffive) \\
  & & & (\ten,\ten,\ffive),(\ten,\fifteen,\ffive),(\ten,\forty,\ffive),(\fifteen,\ten,\ffive),(\fifteen,\fifteen,\ffive), \\
  \multirow{-2}{*}{$\frac{1}{6}$} & \multirow{-2}{*}{$1$} & \multirow{-2}{*}{$-\frac{7}{6}$} & (\fifteen,\forty,\ffive),(\tfive,\forty,\ffive),(\forty,\ten,\ffive),(\forty,\fifteen,\ffive),(\forty,\forty,\ffive) \\
  \rowcolor{blue!10} & & & (\ten,\tenS,\24),(\ten,\fifteenS,\24),(\ten,\tfiveS,\24),(\ten,\fortyS,\24),(\fifteen,\tenS,\24),(\fifteen,\fifteenS,\24),(\fifteen,\fortyS,\24),(\tfive,\tenS,\24), \\
  \rowcolor{blue!10} \multirow{-2}{*}{$-\frac{2}{3}$} & \multirow{-2}{*}{$-\frac{1}{6}$} & \multirow{-2}{*}{$\frac{5}{6}$} & (\tfive,\fifteenS,\24),(\tfive,\tfiveS,\24),(\tfive,\fortyS,\24),(\forty,\tenS,\24),(\forty,\fifteenS,\24),(\forty,\tfiveS,\24),(\forty,\fortyS,\24) \\
  & & & (\ten,\tenS,\24),(\ten,\fifteenS,\24),(\ten,\fortyS,\24),(\fifteen,\tenS,\24),(\fifteen,\fifteenS,\24),(\fifteen,\fortyS,\24), \\
  \multirow{-2}{*}{$\frac{1}{6}$} & \multirow{-2}{*}{$-1$} & \multirow{-2}{*}{$\frac{5}{6}$} & (\tfive,\tfiveS,\24),(\tfive,\fortyS,\24),(\forty,\tenS,\24),(\forty,\fifteenS,\24),(\forty,\tfiveS,\24),(\forty,\fortyS,\24) \\
  \rowcolor{blue!10}  $-\frac{2}{3}$ & $-\frac{5}{6}$ & $\frac{3}{2}$ & (\ten,\24,\fortyS),(\tfive,\24,\tfiveS),(\tfive,\24,\fortyS),(\forty,\24,\tfiveS),(\forty,\24,\fortyS) \\
  $\frac{1}{6}$ & $-\frac{3}{2}$ & $\frac{4}{3}$ & (\ten,\forty,\ffive),(\fifteen,\forty,\ffive),(\forty,\forty,\ffive) \\
  \rowcolor{blue!10} $\frac{1}{6}$ & $-\frac{4}{3}$ & $\frac{7}{6}$ & (\ten,\ffiveS,\ffiveS),(\fifteen,\ffiveS,\ffiveS),(\tfive,\ffiveS,\ffiveS),(\forty,\ffiveS,\ffiveS) \\
      \hline
  \end{tabular}
  \caption{
  \small Hyperons charged under the three individual hypercharge groups and their \555 origin. All \555 representations that involve up to $\ffive$ and $\ffiveS$ of $\GGM$ are included. Other hyperons can be obtained by reordering the hypercharge values or by conjugating the \555 representations.
  \label{tab:hyperons2}
  }
\end{table}
}

\bibliographystyle{JHEP}
\bibliography{Refs_Flavour}

\providecommand{\href}[2]{#2}\begingroup\raggedright\begin{thebibliography}{10}

\bibitem{FernandezNavarro:2023rhv}
M.~Fern\'andez~Navarro and S.~F. King, \emph{{Tri-hypercharge: a separate gauged weak hypercharge for each fermion family as the origin of flavour}}, \href{https://doi.org/10.1007/JHEP08(2023)020}{\emph{JHEP} {\bfseries 08} (2023) 020} [\href{https://arxiv.org/abs/2305.07690}{{\ttfamily 2305.07690}}].

\bibitem{Georgi:1974sy}
H.~Georgi and S.~L. Glashow, \emph{{Unity of All Elementary Particle Forces}}, \href{https://doi.org/10.1103/PhysRevLett.32.438}{\emph{Phys. Rev. Lett.} {\bfseries 32} (1974) 438}.

\bibitem{Salam:1979p}
A.~Salam, \emph{{A gauge appreciation of developments in particle physics}},  in \emph{{\href{https://cds.cern.ch/record/101473/files/C79.06.27_Vol. 2.pdf}{Proceedings of the European Physical Society International Conference on High Energy Physics}}}, (CERN, Geneva, 1979), footnote 41 therein.

\bibitem{Rajpoot:1980ib}
S.~Rajpoot, \emph{{Some Consequences of Extending the SU(5) Gauge Symmetry to the Generation Symmetry SU(5) $e$ X SU(5) $\mu$ X SU(5) $\tau$}}, \href{https://doi.org/10.1103/PhysRevD.24.1890}{\emph{Phys. Rev. D} {\bfseries 24} (1981) 1890}.

\bibitem{Georgi:1981gj}
H.~Georgi, \emph{{Composite/Fundamental Higgs Mesons II: Model Building}}, \href{https://doi.org/10.1016/0550-3213(82)90406-0}{\emph{Nucl. Phys. B} {\bfseries 202} (1982) 397}.

\bibitem{Barbieri:1994cx}
R.~Barbieri, G.~R. Dvali and A.~Strumia, \emph{{Fermion masses and mixings in a flavor symmetric GUT}}, \href{https://doi.org/10.1016/0550-3213(94)00510-L}{\emph{Nucl. Phys. B} {\bfseries 435} (1995) 102} [\href{https://arxiv.org/abs/hep-ph/9407239}{{\ttfamily hep-ph/9407239}}].

\bibitem{Chou:1998pra}
C.-L. Chou, \emph{{Fermion mass hierarchy without flavor symmetry}}, \href{https://doi.org/10.1103/PhysRevD.58.093018}{\emph{Phys. Rev. D} {\bfseries 58} (1998) 093018} [\href{https://arxiv.org/abs/hep-ph/9804325}{{\ttfamily hep-ph/9804325}}].

\bibitem{Asaka:2004ry}
T.~Asaka and Y.~Takanishi, \emph{{Masses and mixing of quarks and leptons in product-group unification}},  \href{https://arxiv.org/abs/hep-ph/0409147}{{\ttfamily hep-ph/0409147}}.

\bibitem{Babu:2007mb}
K.~S. Babu, S.~M. Barr and I.~Gogoladze, \emph{{Family Unification with SO(10)}}, \href{https://doi.org/10.1016/j.physletb.2008.01.057}{\emph{Phys. Lett. B} {\bfseries 661} (2008) 124} [\href{https://arxiv.org/abs/0709.3491}{{\ttfamily 0709.3491}}].

\bibitem{Li:1981nk}
X.~Li and E.~Ma, \emph{{Gauge Model of Generation Nonuniversality}}, \href{https://doi.org/10.1103/PhysRevLett.47.1788}{\emph{Phys. Rev. Lett.} {\bfseries 47} (1981) 1788}.

\bibitem{Ma:1987ds}
E.~Ma, X.~Li and S.~F. Tuan, \emph{{Gauge Model of Generation Nonuniversality Revisited}}, \href{https://doi.org/10.1103/PhysRevLett.60.495}{\emph{Phys. Rev. Lett.} {\bfseries 60} (1988) 495}.

\bibitem{Ma:1988dn}
E.~Ma and D.~Ng, \emph{{Gauge and Higgs Bosons in a Model of Generation Nonuniversality}}, \href{https://doi.org/10.1103/PhysRevD.38.304}{\emph{Phys. Rev. D} {\bfseries 38} (1988) 304}.

\bibitem{Li:1992fi}
X.-y. Li and E.~Ma, \emph{{Gauge model of generation nonuniversality reexamined}}, \href{https://doi.org/10.1088/0954-3899/19/9/006}{\emph{J. Phys. G} {\bfseries 19} (1993) 1265} [\href{https://arxiv.org/abs/hep-ph/9208210}{{\ttfamily hep-ph/9208210}}].

\bibitem{Muller:1996dj}
D.~J. Muller and S.~Nandi, \emph{{Top flavor: A Separate SU(2) for the third family}}, \href{https://doi.org/10.1016/0370-2693(96)00745-9}{\emph{Phys. Lett. B} {\bfseries 383} (1996) 345} [\href{https://arxiv.org/abs/hep-ph/9602390}{{\ttfamily hep-ph/9602390}}].

\bibitem{Chiang:2009kb}
C.-W. Chiang, N.~G. Deshpande, X.-G. He and J.~Jiang, \emph{{The Family $SU(2)_l$ x $SU(2)_h$ x $U(1)$ Model}}, \href{https://doi.org/10.1103/PhysRevD.81.015006}{\emph{Phys. Rev. D} {\bfseries 81} (2010) 015006} [\href{https://arxiv.org/abs/0911.1480}{{\ttfamily 0911.1480}}].

\bibitem{Carone:1995ge}
C.~D. Carone and H.~Murayama, \emph{{Third family flavor physics in an $SU(3) ^{3}$ x SU(2) -L x U(1) -Y model}}, \href{https://doi.org/10.1103/PhysRevD.52.4159}{\emph{Phys. Rev. D} {\bfseries 52} (1995) 4159} [\href{https://arxiv.org/abs/hep-ph/9504393}{{\ttfamily hep-ph/9504393}}].

\bibitem{Bordone:2017bld}
M.~Bordone, C.~Cornella, J.~Fuentes-Martin and G.~Isidori, \emph{{A three-site gauge model for flavor hierarchies and flavor anomalies}}, \href{https://doi.org/10.1016/j.physletb.2018.02.011}{\emph{Phys. Lett. B} {\bfseries 779} (2018) 317} [\href{https://arxiv.org/abs/1712.01368}{{\ttfamily 1712.01368}}].

\bibitem{Greljo:2018tuh}
A.~Greljo and B.~A. Stefanek, \emph{{Third family quark\textendash{}lepton unification at the TeV scale}}, \href{https://doi.org/10.1016/j.physletb.2018.05.033}{\emph{Phys. Lett. B} {\bfseries 782} (2018) 131} [\href{https://arxiv.org/abs/1802.04274}{{\ttfamily 1802.04274}}].

\bibitem{Allwicher:2020esa}
L.~Allwicher, G.~Isidori and A.~E. Thomsen, \emph{{Stability of the Higgs Sector in a Flavor-Inspired Multi-Scale Model}}, \href{https://doi.org/10.1007/JHEP01(2021)191}{\emph{JHEP} {\bfseries 01} (2021) 191} [\href{https://arxiv.org/abs/2011.01946}{{\ttfamily 2011.01946}}].

\bibitem{Fuentes-Martin:2020pww}
J.~Fuentes-Martin, G.~Isidori, J.~Pag\`es and B.~A. Stefanek, \emph{{Flavor non-universal Pati-Salam unification and neutrino masses}}, \href{https://doi.org/10.1016/j.physletb.2021.136484}{\emph{Phys. Lett. B} {\bfseries 820} (2021) 136484} [\href{https://arxiv.org/abs/2012.10492}{{\ttfamily 2012.10492}}].

\bibitem{Fuentes-Martin:2022xnb}
J.~Fuentes-Martin, G.~Isidori, J.~M. Lizana, N.~Selimovic and B.~A. Stefanek, \emph{{Flavor hierarchies, flavor anomalies, and Higgs mass from a warped extra dimension}}, \href{https://doi.org/10.1016/j.physletb.2022.137382}{\emph{Phys. Lett. B} {\bfseries 834} (2022) 137382} [\href{https://arxiv.org/abs/2203.01952}{{\ttfamily 2203.01952}}].

\bibitem{Davighi:2022bqf}
J.~Davighi, G.~Isidori and M.~Pesut, \emph{{Electroweak-flavour and quark-lepton unification: a family non-universal path}}, \href{https://doi.org/10.1007/JHEP04(2023)030}{\emph{JHEP} {\bfseries 04} (2023) 030} [\href{https://arxiv.org/abs/2212.06163}{{\ttfamily 2212.06163}}].

\bibitem{Davighi:2022fer}
J.~Davighi and J.~Tooby-Smith, \emph{{Electroweak flavour unification}}, \href{https://doi.org/10.1007/JHEP09(2022)193}{\emph{JHEP} {\bfseries 09} (2022) 193} [\href{https://arxiv.org/abs/2201.07245}{{\ttfamily 2201.07245}}].

\bibitem{Davighi:2023iks}
J.~Davighi and G.~Isidori, \emph{{Non-universal gauge interactions addressing the inescapable link between Higgs and flavour}}, \href{https://doi.org/10.1007/JHEP07(2023)147}{\emph{JHEP} {\bfseries 07} (2023) 147} [\href{https://arxiv.org/abs/2303.01520}{{\ttfamily 2303.01520}}].

\bibitem{Glashow:1984gc}
S.~L. Glashow, \emph{{Trinification of All Elementary Particle Forces}},  in \emph{{Fifth Workshop on Grand Unification}}, 7, 1984.

\bibitem{Malkawi:1996fs}
E.~Malkawi, T.~M.~P. Tait and C.~P. Yuan, \emph{{A Model of strong flavor dynamics for the top quark}}, \href{https://doi.org/10.1016/0370-2693(96)00859-3}{\emph{Phys. Lett. B} {\bfseries 385} (1996) 304} [\href{https://arxiv.org/abs/hep-ph/9603349}{{\ttfamily hep-ph/9603349}}].

\bibitem{Shu:2006mm}
J.~Shu, T.~M.~P. Tait and C.~E.~M. Wagner, \emph{{Baryogenesis from an Earlier Phase Transition}}, \href{https://doi.org/10.1103/PhysRevD.75.063510}{\emph{Phys. Rev. D} {\bfseries 75} (2007) 063510} [\href{https://arxiv.org/abs/hep-ph/0610375}{{\ttfamily hep-ph/0610375}}].

\bibitem{Barbieri:2011ci}
R.~Barbieri, G.~Isidori, J.~Jones-Perez, P.~Lodone and D.~M. Straub, \emph{{$U(2)$ and Minimal Flavour Violation in Supersymmetry}}, \href{https://doi.org/10.1140/epjc/s10052-011-1725-z}{\emph{Eur. Phys. J. C} {\bfseries 71} (2011) 1725} [\href{https://arxiv.org/abs/1105.2296}{{\ttfamily 1105.2296}}].

\bibitem{Allwicher:2023shc}
L.~Allwicher, C.~Cornella, B.~A. Stefanek and G.~Isidori, \emph{{New Physics in the Third Generation: A Comprehensive SMEFT Analysis and Future Prospects}},  \href{https://arxiv.org/abs/2311.00020}{{\ttfamily 2311.00020}}.

\bibitem{Davighi:2023evx}
J.~Davighi and B.~A. Stefanek, \emph{{Deconstructed Hypercharge: A Natural Model of Flavour}},  \href{https://arxiv.org/abs/2305.16280}{{\ttfamily 2305.16280}}.

\bibitem{Georgi:1979df}
H.~Georgi and C.~Jarlskog, \emph{{A New Lepton - Quark Mass Relation in a Unified Theory}}, \href{https://doi.org/10.1016/0370-2693(79)90842-6}{\emph{Phys. Lett. B} {\bfseries 86} (1979) 297}.

\bibitem{Ferretti:2006df}
L.~Ferretti, S.~F. King and A.~Romanino, \emph{{Flavour from accidental symmetries}}, \href{https://doi.org/10.1088/1126-6708/2006/11/078}{\emph{JHEP} {\bfseries 11} (2006) 078} [\href{https://arxiv.org/abs/hep-ph/0609047}{{\ttfamily hep-ph/0609047}}].

\bibitem{UTfit:2007eik}
{\scshape UTfit} collaboration, M.~Bona et~al., \emph{{Model-independent constraints on $\Delta F=2$ operators and the scale of new physics}}, \href{https://doi.org/10.1088/1126-6708/2008/03/049}{\emph{JHEP} {\bfseries 03} (2008) 049} [\href{https://arxiv.org/abs/0707.0636}{{\ttfamily 0707.0636}}].

\bibitem{Isidori:2014rba}
G.~Isidori and F.~Teubert, \emph{{Status of indirect searches for New Physics with heavy flavour decays after the initial LHC run}}, \href{https://doi.org/10.1140/epjp/i2014-14040-4}{\emph{Eur. Phys. J. Plus} {\bfseries 129} (2014) 40} [\href{https://arxiv.org/abs/1402.2844}{{\ttfamily 1402.2844}}].

\bibitem{Fonseca:2020vke}
R.~M. Fonseca, \emph{{GroupMath: A Mathematica package for group theory calculations}}, \href{https://doi.org/10.1016/j.cpc.2021.108085}{\emph{Comput. Phys. Commun.} {\bfseries 267} (2021) 108085} [\href{https://arxiv.org/abs/2011.01764}{{\ttfamily 2011.01764}}].

\bibitem{deSalas:2020pgw}
P.~F. de~Salas, D.~V. Forero, S.~Gariazzo, P.~Mart\'\i{}nez-Mirav\'e, O.~Mena, C.~A. Ternes et~al., \emph{{2020 global reassessment of the neutrino oscillation picture}}, \href{https://doi.org/10.1007/JHEP02(2021)071}{\emph{JHEP} {\bfseries 02} (2021) 071} [\href{https://arxiv.org/abs/2006.11237}{{\ttfamily 2006.11237}}].

\bibitem{Gonzalez-Garcia:2021dve}
M.~C. Gonzalez-Garcia, M.~Maltoni and T.~Schwetz, \emph{{NuFIT: Three-Flavour Global Analyses of Neutrino Oscillation Experiments}}, \href{https://doi.org/10.3390/universe7120459}{\emph{Universe} {\bfseries 7} (2021) 459} [\href{https://arxiv.org/abs/2111.03086}{{\ttfamily 2111.03086}}].

\bibitem{Machacek:1983tz}
M.~E. Machacek and M.~T. Vaughn, \emph{{Two Loop Renormalization Group Equations in a General Quantum Field Theory. 1. Wave Function Renormalization}}, \href{https://doi.org/10.1016/0550-3213(83)90610-7}{\emph{Nucl. Phys. B} {\bfseries 222} (1983) 83}.

\bibitem{Super-Kamiokande:2020wjk}
{\scshape Super-Kamiokande} collaboration, A.~Takenaka et~al., \emph{{Search for proton decay via $p\to e^+\pi^0$ and $p\to \mu^+\pi^0$ with an enlarged fiducial volume in Super-Kamiokande I-IV}}, \href{https://doi.org/10.1103/PhysRevD.102.112011}{\emph{Phys. Rev. D} {\bfseries 102} (2020) 112011} [\href{https://arxiv.org/abs/2010.16098}{{\ttfamily 2010.16098}}].

\bibitem{Bhattiprolu:2022xhm}
P.~N. Bhattiprolu, S.~P. Martin and J.~D. Wells, \emph{{Statistical significances and projections for proton decay experiments}}, \href{https://doi.org/10.1103/PhysRevD.107.055016}{\emph{Phys. Rev. D} {\bfseries 107} (2023) 055016} [\href{https://arxiv.org/abs/2210.07735}{{\ttfamily 2210.07735}}].

\bibitem{Nath:2006ut}
P.~Nath and P.~Fileviez~Perez, \emph{{Proton stability in grand unified theories, in strings and in branes}}, \href{https://doi.org/10.1016/j.physrep.2007.02.010}{\emph{Phys. Rept.} {\bfseries 441} (2007) 191} [\href{https://arxiv.org/abs/hep-ph/0601023}{{\ttfamily hep-ph/0601023}}].

\bibitem{Chakrabortty:2019fov}
J.~Chakrabortty, R.~Maji and S.~F. King, \emph{{Unification, Proton Decay and Topological Defects in non-SUSY GUTs with Thresholds}}, \href{https://doi.org/10.1103/PhysRevD.99.095008}{\emph{Phys. Rev. D} {\bfseries 99} (2019) 095008} [\href{https://arxiv.org/abs/1901.05867}{{\ttfamily 1901.05867}}].

\bibitem{Aoki:2017puj}
Y.~Aoki, T.~Izubuchi, E.~Shintani and A.~Soni, \emph{{Improved lattice computation of proton decay matrix elements}}, \href{https://doi.org/10.1103/PhysRevD.96.014506}{\emph{Phys. Rev. D} {\bfseries 96} (2017) 014506} [\href{https://arxiv.org/abs/1705.01338}{{\ttfamily 1705.01338}}].

\end{thebibliography}\endgroup

\end{document}